\documentclass[structabstract]{aa}  
\usepackage{natbib}
\usepackage{graphicx}
\usepackage{txfonts}
\newcommand{\dbscan}{DBSCAN}
\newcommand{\gray}{$\gamma$-ray}
\newcommand{\gdbscan}{$\gamma$-ray DBSCAN}

\newcommand{\FLAT}{{\it Fermi}-LAT}
\newcommand{\Eps}{$\varepsilon$}

\begin{document}

\title{\gray~\dbscan: a clustering algorithm applied to \FLAT~ \gray~ data.}

   \subtitle{I: Detection performances  with real and simulated  data}

   \author{
        A. Tramacere  \inst{1}
        \and
        C. Vecchio  \inst{2}
}       
          
   \institute{
        ISDC, University of Geneva, Chemin d'Ecogia 16, Versoix, CH-1290, Switzerland \email{andrea.tramacere@unige.ch} \label{ISDC}
        \and
        Politecnico di Milano - Piazza L. da Vinci, 32 - 20133 Milano, Italy            \label{Politecnico}              
            }
   \date{Received; accepted}

  \abstract
  {The Density Based Spatial Clustering of Applications with Noise 
  (\dbscan) is  a topometric algorithm used to cluster spatial data that are affected by background noise. For the first
  time, we propose the use of this method for the detection of sources  in \gray~ astrophysical images obtained from
  the \FLAT~ data, where each point corresponds to {  the arrival direction of a photon}.}
  {We investigate the detection performance of the  \gdbscan~ in terms of detection efficiency and 
  rejection of spurious clusters. }
  {We use a parametric approach, exploring a large volume of the \gdbscan~ parameter space. By means  
  of simulated data we statistically characterize the \gdbscan, finding signatures that differentiate  
  purely random fields, from fields with
  sources. We define a significance level for the detected clusters, and we successfully test this  
  significance with our simulated
  data. We apply the method to real data, and we find an excellent agreement with the results  
  obtained with simulated data.  }
  {{   
  We find that the  \gdbscan~ can be successfully used in the detection of clusters in \gray~ data.    
  The significance returned by our algorithm is strongly correlated with that provided by the Maximum  
  Likelihood analysis with standard \FLAT~ software, and can be used to safely remove spurious  
  clusters. The positional accuracy of the reconstructed cluster centroid compares to that  returned by standard Maximum  
  Likelihood analysis, allowing to look  for astrophysical counterparts in narrow regions,  minimizing the 
  chance probability in the counterpart association.
  }}
  {{     
  We find that \gdbscan~ is a powerful tool in the detection of clusters in \gray~ data, this method can be  
  used both to look for	point-like sources, and extended sources, and can be potentially applied to any 
  astrophysical field related with detection of clusters in data. In a companion paper we will  
  present the application of the \gdbscan~ to the full	\FLAT~ sky, discussing  the potentiality in the  discovery of new sources.
  }}

\titlerunning{ \gdbscan~ application to \FLAT~ \gray~ data.}
\authorrunning{A. Tramacere and C. Vecchio}

   \maketitle
\section{Introduction}
Modern \gray~  telescopes operating at energies above the MeV window, provide
event-resolved observational data. Each event (after the reconstruction process)
is typically described by a tuple (i.e. an ordered list of elements) storing sky 
coordinates, arrival time, and energy. 
Detection of discrete sources (either point-like or extended) is
performed using various methods. Given the discrete topological nature of \gray~
images, methods based on cluster search, like the {\it  { Minimum} Spanning Three}
(MST) \citep{MSTI,MSTII} have successfully been used. One of the main advantages of topometric
methods, { compared to methods using the spatial binning},  is to minimize 
the impact of the poor energy-dependent Point Spread Function (PSF), typical of
\gray~ telescopes, preserving the spatial information of each event. Moreover,
these methods are able to detect sources compounded by a small amount of events,
but they need to be fine tuned to take into account properly the background. The
problem of background rejection is the most penalizing feature of topometric
methods, for this reason in this paper, for the first time, we present a 
method based on the \dbscan~ algorithm \citep{Ester96DBScan}. 
The \dbscan~ is a topometric algorithm used to
cluster spatial data that are affected by background noise. Compared to other
topometric methods, it has the advantage to embed inside the algorithm itself the
discrimination between signal (cluster) and background (noise), according to the
local density of events within a typical scanning brush { i.e. within 
a given scanning area}.

{ The aim of the present paper is to show the potentialities of the method,
and  its statistical characterization when applied to astrophysical \gray~ data.} 
We apply this method to the detection of point-like sources in the \FLAT~ data. 
We explore a large volume of the \gdbscan~ parameter space, by means  of simulated 
data, and we provide a statistical characterization the \gdbscan, finding signatures 
that differentiate  purely random fields, from fields with sources. We define a 
significance level for the detected clusters, and we successfully test this significance 
with our simulated data. We apply the method to real \FLAT~ \gray~ data, and we 
find an excellent agreement with the results  obtained with simulated data.

In a companion paper \citep{dbscan_inprep}, we will apply the method
to the	 \FLAT~ sky, {investigating specific issues related to the \FLAT~ response 
functions}, showing the potentiality for the discovery of new sources,
in particular of small clusters located at high galactic latitude, or   
clusters on the galactic plane, affected by a strong background.

The paper is organized as follows. In Sec. \ref{sec:dbscan} we describe the logic of
the \dbscan~ method, and we present the algorithm implemented to analyse \gray~
data, the \gdbscan.
In Sec. \ref{sec:caveat} we discuss some caveats regarding the
application of the \gdbscan~ algorithm to \gray~ data. 
In Sec. \ref{sec:stats_sim} we study the statistical properties of the
\gdbscan~ detection, using a simulated test field with only noise,
and five simulated test fields with noise plus point-like sources.
In Sec. \ref{sec:det_sim}  we evaluate the detection  performance of the method 
in terms of positional accuracy, cluster reconstruction,  and rejection of spurious clusters.
In Sec. \ref{sec:signif_sim} we investigate the significance of the clusters,
and describe our algorithmic implementation.
In Section \ref{sec:realdata} we finally use our method with real \FLAT~ data,
investigating the detection performance, and comparing the \gdbscan~ clusters 
significance, to that returned by the Maximum Likelihood method with
standard \FLAT~ software \footnote{http://fermi.gsfc.nasa.gov/ssc/data/analysis/scitools/overview.html}.
In Section \ref{sec:conclusions}, we present our conclusions, and we discuss
future developments and applications.

\begin{figure*}
\centering
\begin{tabular}{l} 
	\includegraphics[width=18cm]{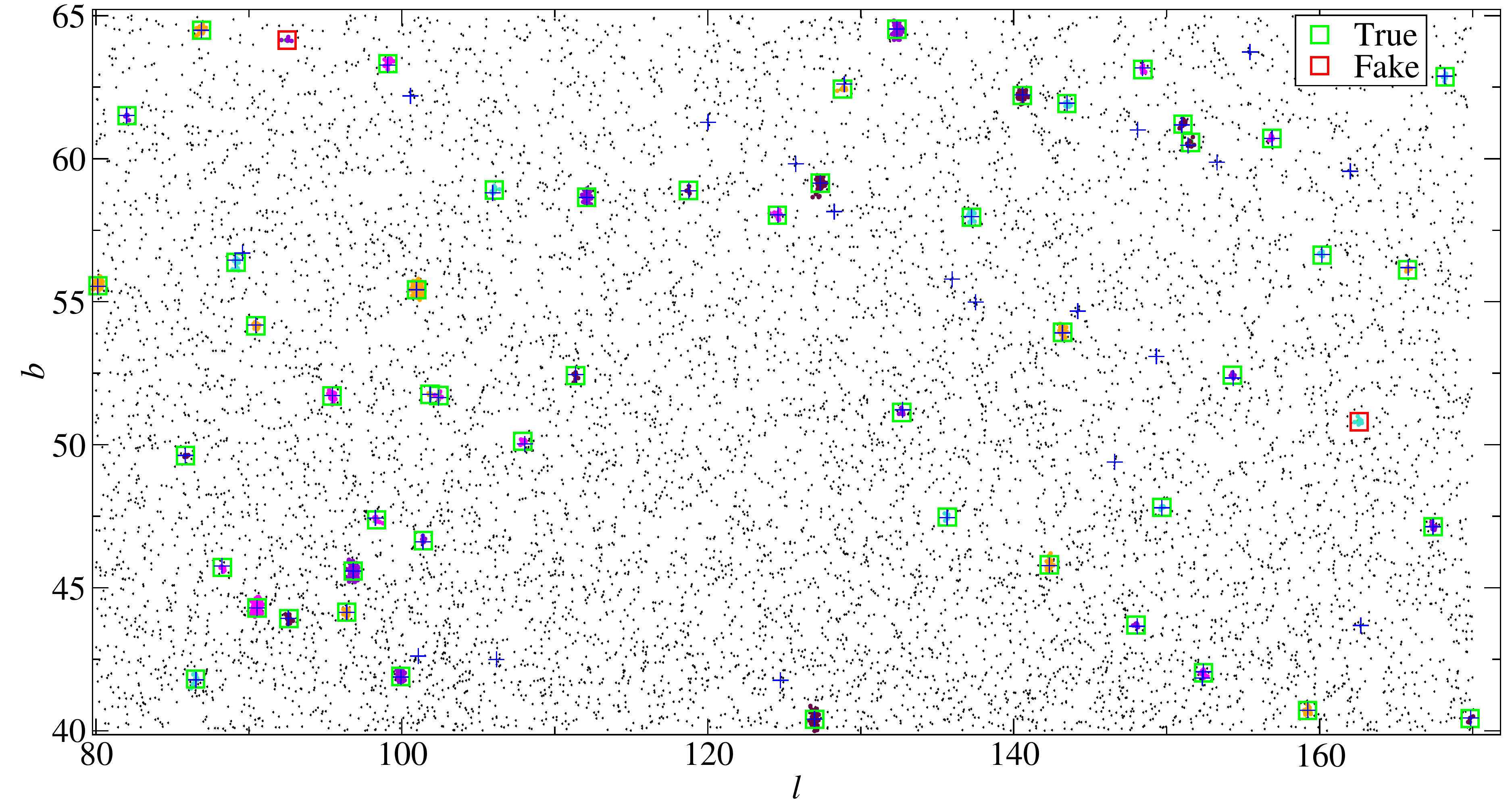}\\
\end{tabular}
\caption{ Photon map for the {\it sky} test field 1, 
with the result of the \gdbscan~ detection for $K = 5$ and $\varepsilon=0.17$ deg.
The blue crosses refer to the simulated sources, the green boxes  
to  51 detected {\it true} clusters, and the red boxes to the 2 {\it fake} ones. 
The black dots represent the background events, the remaining colors indicate
cluster events.}
\label{fig:sim_detection}
\end{figure*}
\section{The \gray~ \dbscan~ algorithm}
\label{sec:dbscan} 
The \dbscan~ \citep{Ester96DBScan} is a topometric algorithm used  to cluster
spatial data that are affected by background noise.  Some modifications have
been developed to adapt the original \dbscan~ algorithm to our study. Our
algorithm is mainly built upon the following criteria: 
\begin{enumerate}
\item{} given a list of photons $D$, where each element $p_i$ is  a  tuple storing  
positional sky coordinates, let $\rho(p_k,p_l)$ be the angular distance between 
two photons $p_k$ and $p_l$.
\item{} We iterate over the full photon list $D$. A seed cluster $C_m^*$ is built 
when a minimum number of photons $K+1$ is enclosed within a circle of radius \Eps~
centered on $p_i$ 
\item{} For each photon  $p_l \in C^*_m$,  we build the photon list $C_m^+$ 
by collecting all the photons $p_k$ respecting the condition:  
$\rho(p_l,p_k)<\varepsilon$, and $p_k \notin C^*_m$. 
\item For each photon  $p_j \in C_m^+$, if the number of photons enclosed within a circle of  
radius \Eps~ centered on $p_j$ is $\leq K$ and $p_j \notin C^*_m$, then $p_j$ will be attached to the 
{ final photon list of the cluster   without a recursive search for further neighbours}, 
these points are defined  {\it density-reachable}.
\item For each photon  $p_j \in C_m^+$, if the number of photons enclosed within a circle of 
radius \Eps~ centered on $p_j$ is $>K$ and $p_j \notin C^*_m$, $p_j$ is attached to the $C_m^*$, and  
and step { 3} is repeated recursively.
\item{} { When both conditions at step 4 and 5 are false, the cluster $C_m$ is built by joining
the {\it density-reachable} events to those in the $C^*_m$ and in the  $C_m^+$  lists.}
{ \item{} The process starts again from step 1 searching for new clusters, skipping all the
events already flagged as {\it noise} or {\it clusters}, until all the
events in $D$ are flagged as {\it cluster}, or {\it noise}, or
{\it density-reachable}, events. }
\item{} At the end of the process the full photon list will be partitioned as 
follows: 
\begin{eqnarray}
D &=& D_{cls} \cup D_{noise}= (p_i \in \cup_m C_m) \cup (p_i \notin  \cup_m C_m) \\
\emptyset&=&D_{cls} \cap D_{noise} \nonumber
\end{eqnarray}
\end{enumerate}
In this way high densely populated areas are classified as clusters (sources),
conversely  low densely populated areas  are classified as noise (background). 
The recursive call  of step 3, is not implemented in the original
\dbscan~ algorithm, and { represents} a novelty. This new feature, allows to reconstruct
clusters with a size significantly larger than the \Eps~ radius, making rare the  
possibility to fragment a single clusters in small satellite clusters. Moreover, allows
the possibility to reconstruct extended { structures}, in particular extended sources, or
filamentary structures in the background.

After the clustering process, each photon in $D$ will be described by a tuple, storing:
{ 
the photon position (both in galactic  and celestial  coordinates),
the photon class type ({\it noise} or {\it cluster}),
and the ID of the cluster the photon belongs to. 
Each cluster $C_m$,  will be described by a tuple storing 
the position of the centroid  with his positional error,   the ellipse  
of the cluster containment, the cluster effective radius ($r_eff$),
and number of photons in the cluster ($N_p$).
The ellipse of the cluster containment, is defined by major and 
minor semi-axis ($\sigma_x$  and $\sigma_y$, respectively), and the inclination angle
($\sigma_{alpha}$) of the major semi-axis w.r.t. the latitudinal coordinate. ($b$ or $DEC$). 
To evaluate the ellipse axis we use the Principal Component Analysis method (PCA) \citep{Jolliffe1986}. This method uses the eigenvalue decomposition of the 
covariance matrix of the  the two position arrays $\mathbf x$, and $\mathbf y$. 
By definition, the square root of the first eigenvalue will correspond 
to $\sigma_x$, and the second to $\sigma_y$.
The axes represent the two orthogonal directions of maximum variance of the cluster.
The effective radius is defined as $r_{eff}=\sqrt{\sigma_x^2 + \sigma_y^2}$.
To find the centroid of the cluster and its uncertainty, we use a weighted average of the 
position of each photon in $C_m$, as follows:
\begin{itemize}
\item{} we define the first order centroid ($C_{ave}$) as the average of the 
position of each cluster photon: $C_{ave}=(<\mathbf x>,<\mathbf y>)$.
\item{} We define the weight array, according to the distance between
$p_k \in C_m$ and $C_{ave}$: $w_k=1/\rho(p_k,C_{ave})$.
\item {} The cluster centroid $C_{ctr}$ will result from average of the position  of
each cluster point weighted by $w_k$.
\item {} The centroid position uncertainty ($pos_{err}$) is determined by propagating the  
error on the weighted average of $C_{ctr}$.  
We have numerically verified that  $pos_{err}$ corresponds to a $\approx 95\%$ 
positional uncertainty. 

\end{itemize}

\begin{figure}[!h]
\includegraphics[width=9cm]{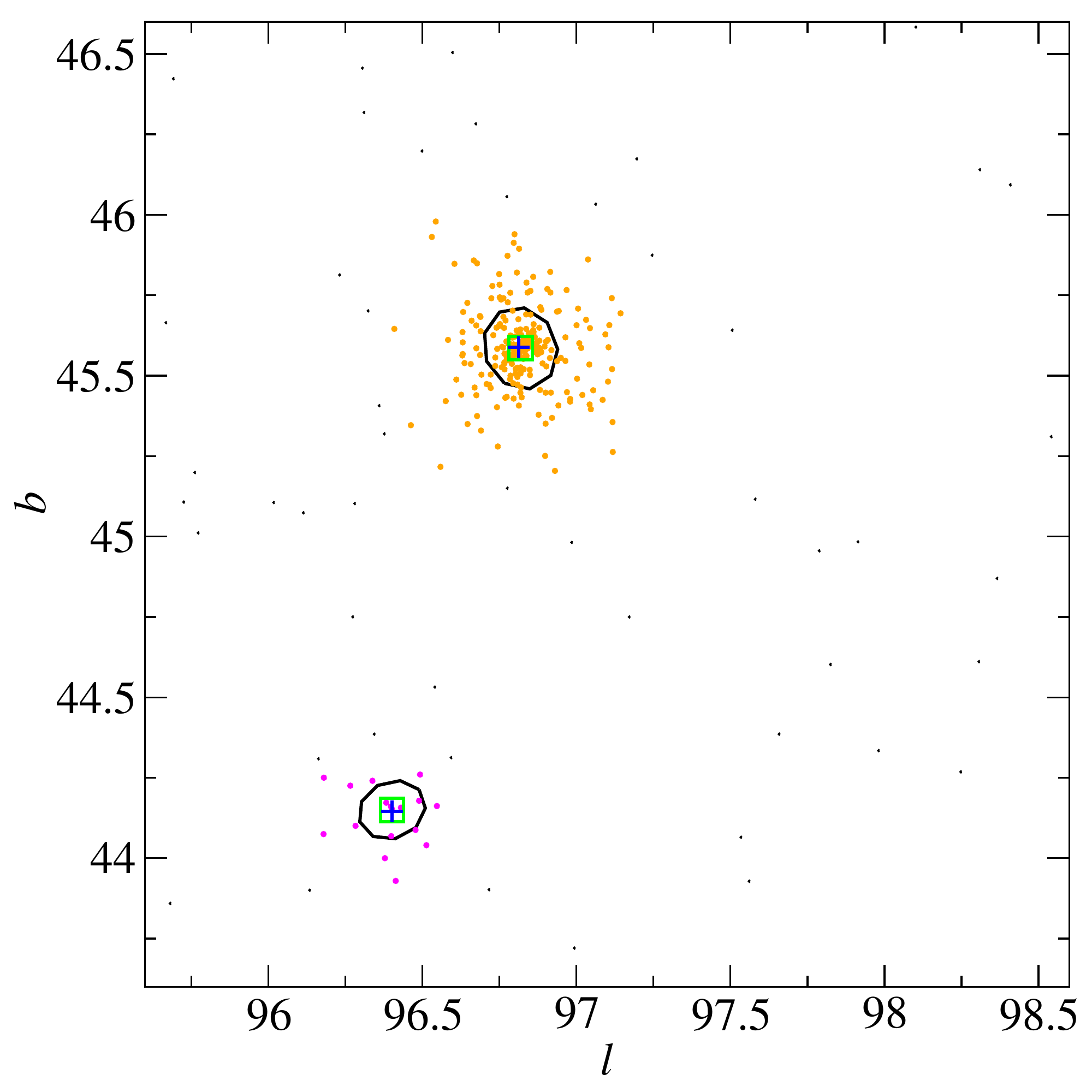}
\caption{
A close-up of two {\it true} clusters reported in Fig. \ref{fig:sim_detection}. 
The  ellipses correspond to the ellipse of  the cluster containment. 
The purple and orange points represent the cluster points, the black dots represent the background
events, the blue crosses the position of the simulated sources, and green boxes the position of 
the clusters centroid
}
\label{fig:closeup}
\end{figure}

\section{Caveat on the application to \gray~ data }
\label{sec:caveat}
The application of clustering methods, such as the \gdbscan, leads to deal
with practical difficulties, related mostly to the instrument PSF, and
to gradient and/or structures in the background. In order to deal with these 
issues, without biasing the detection results, it's recommended to apply some
criteria that we discuss in the following.

 As first, we comment on the PSF impact. The PSF imposes a limit on the capability
of an instrument to resolve sources separated by a distance smaller then the PSF
size. Sources with sizes below the PSF { } classified as point-like, { otherwise} are classified as extended. A further complication is that the PSF
often depends on the energy; in the case of \FLAT,  the $68\%$ containment angle of the 
reconstructed incoming photon direction, for normal incidence photons, has a typical 
size of a couple of degrees at 100 MeV \citep{LATCalib}, and scales down to few tenths of 
degree above the GeV energies 
\footnote{http://www.slac.stanford.edu/exp/glast/groups/canda/\\	lat\_Performance.html}. 
The size of the PSF is strongly connected to the size \Eps~ of the 
\gdbscan~ scanning brush. Indeed, if \Eps~ is much smaller than the PSF size, it 
might occur the risk to loose clusters characterized by small $N_p$, or to fragment a 
cluster with large $N_p$ in smaller fake {\it satellite} clusters.  
We stress that the formation of {\it satellite} clusters is a very rare event, 
thanks to our recursive  DBSCAN implementation, that is explained in Sec. \ref{sec:dbscan}. 
On the contrary, if \Eps~ is much larger w.r.t the PSF, it is likely to build extended 
{ clusters} contaminated by the background, or by  close sources. \\
{
A careful and self-consistent analysis of the effects of the energy dependence
of the PSF, and in general of  issues related to the \FLAT~ response function, 
is beyond the scope of this paper, where we focus mostly on a statistical 
characterization  of the method.
These subjects will be investigated in the companion paper \citep{dbscan_inprep}.
}

A second relevant issue, is the inhomogeneity of the background, that affects
both the choice of \Eps~ and $K$. If the background is homogeneous over the
entire field, the optimal choice of a single pair of values of \Eps~ and $K$,
guarantees a safe rejection of the background. Indeed, values of \Eps~ and $K$,
such that the average density of photons within \Eps~ is significantly larger
the the average density of the background photons, make rare the chance to grow
a cluster from a background fluctuation. Unfortunately, the \gray~ sky shows
strong gradients of background, in particular at low galactic latitudes. To
solve this issue, one could think to adapt the value of \Eps~ and $K$ according
to a local value of the background photon density. Since \Eps~ has a strong
constraint imposed by the PSF, one should tune mostly the value of $K$. The
drawback is that as we increase the value of $K$ to compensate for the
background, we decrease the capability to detect cluster with small $N_p$. To
{ overcome} this difficulty, we adopt an alternative solution. We use a unique pair
of values of \Eps~ and $K$, for each field, where \Eps~ is mostly constrained by
the PSF, and $K$, by the field average background, ad we take into account the
background inhomogeneities by defining a significance level of the cluster,
according to the signal to noise ratio \citep{LiMa1983}, evaluated  from the
local background. This is explained in detail in Section \ref{sec:signif_sim}. The
capability to reject clusters according to a low significance level, allows to
relax the constrain on  \Eps~ and $K$, increasing the number of clusters 
detected, hence increasing the detection ratio, and at the same time allows to
reject spurious sources, due to the significance threshold. 
Anyhow, to avoid that the background is { so high}, that the fluctuations 
in the background events, can lead to densities comparable to those
of weak sources, it's recommended to apply a cut in energy, to make
this possibility rare. 
In order   to optimize the ratio between background and clusters events, in the 
following we use a threshold energy of 3 GeV, that mitigates the possible bias
due to the background fluctuations.

\begin{figure*}
\centering
\begin{tabular}{ll} 
\includegraphics[width=9cm]{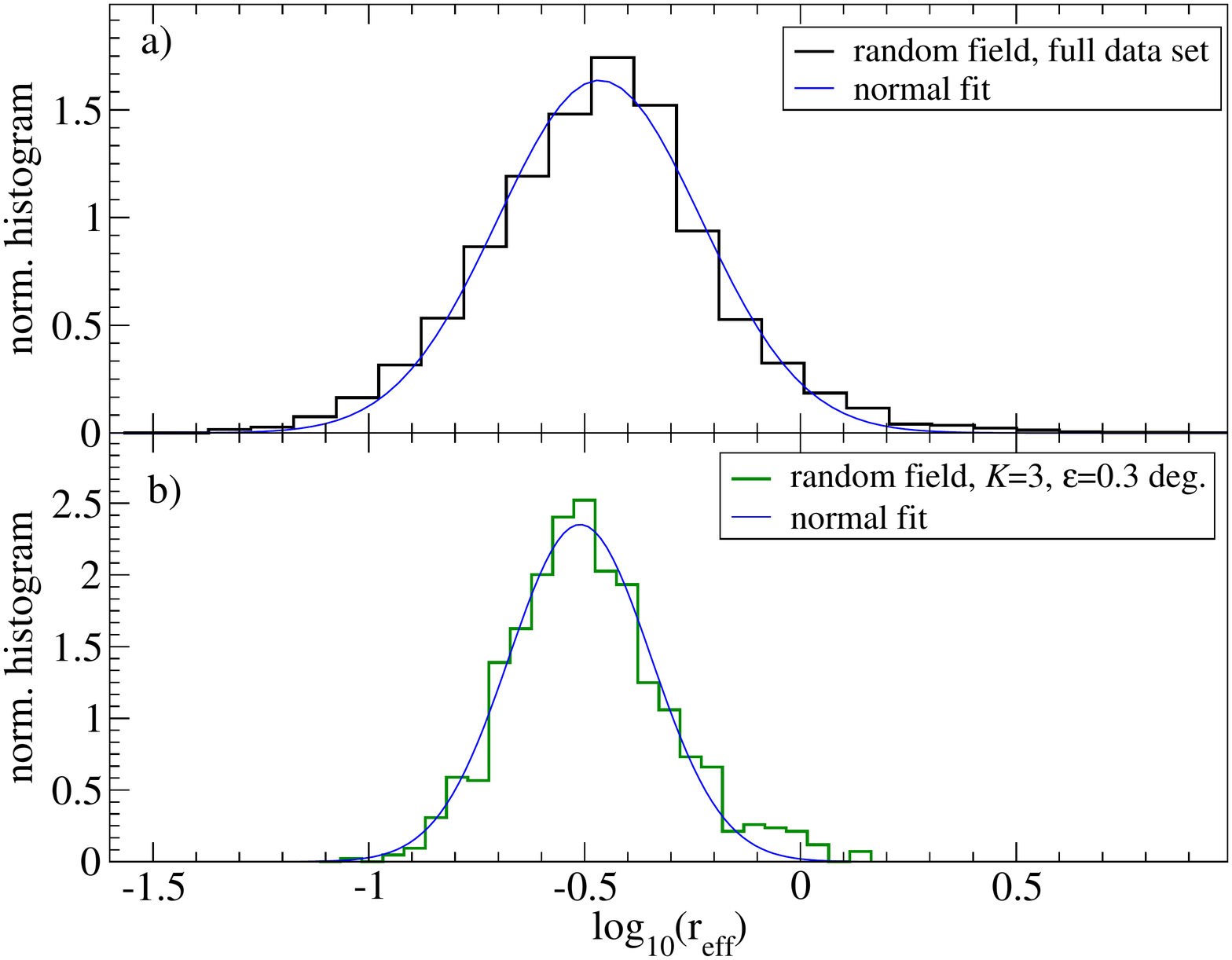}&
\includegraphics[width=9cm]{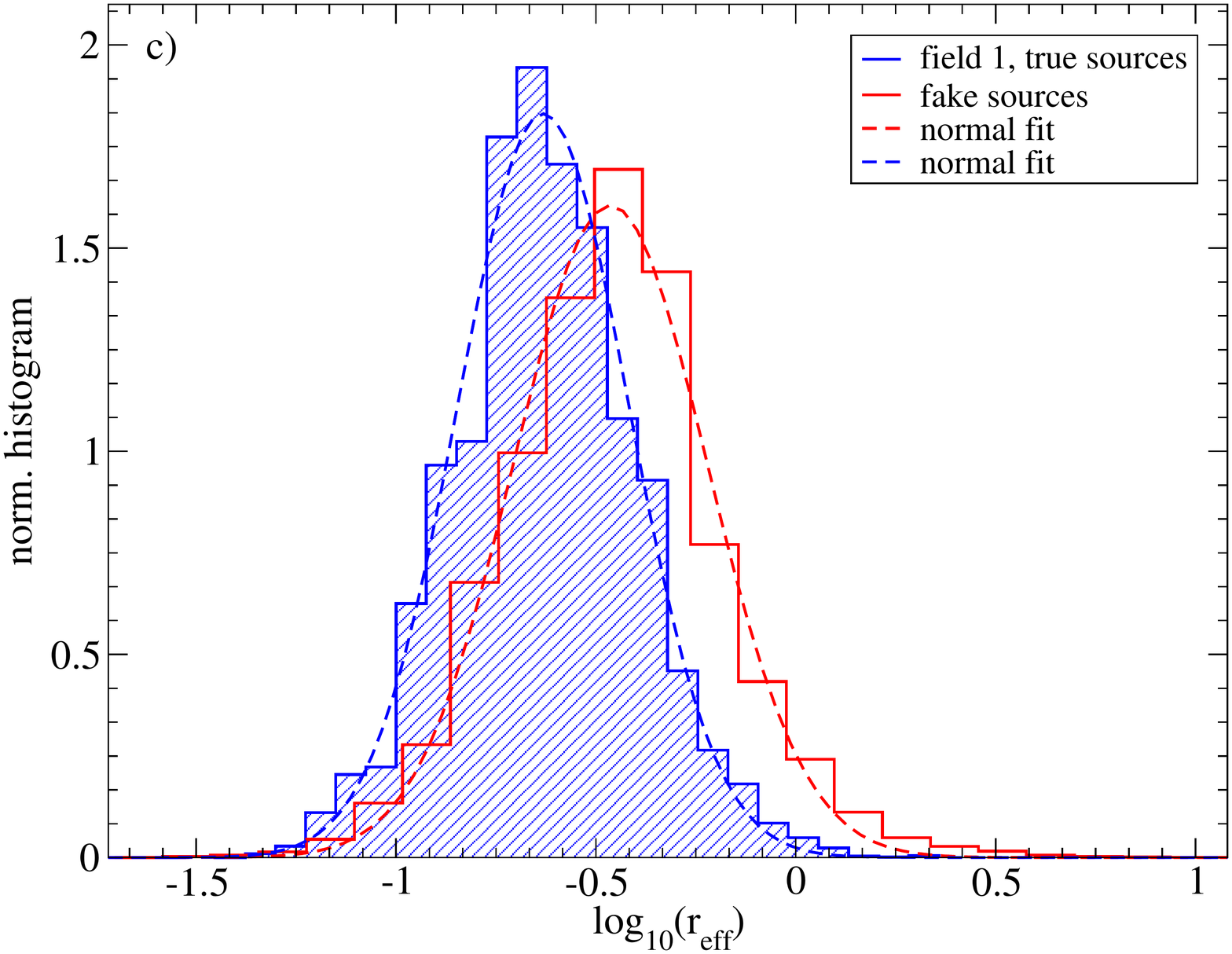}\\
\end{tabular}
\caption{{\it Panel a:} distribution of the values of $\log_{10} r_{eff}$ for 
the {\it random} field case, for  the full parameter space (black line) and fit 
by means  of Gaussian distribution (blue line). {\it Panel b:} 
the same as in the top panel, for the case of $K=3$ and $\varepsilon=0.3$ deg. 
{\it Panel c:}  distribution of $\log_{10} r_{eff} $ for the case of
the {\it sky} test field 1, for {\it fake} clusters (red solid line), 
and {\it true} clusters (blue solid line, hatched histogram). 
{ the  dashed lines represent a Gaussian best fit.}
}
\label{fig:r_68}
\end{figure*}

\begin{figure*}
\centering
\begin{tabular}{ll} 
\includegraphics[width=9cm]{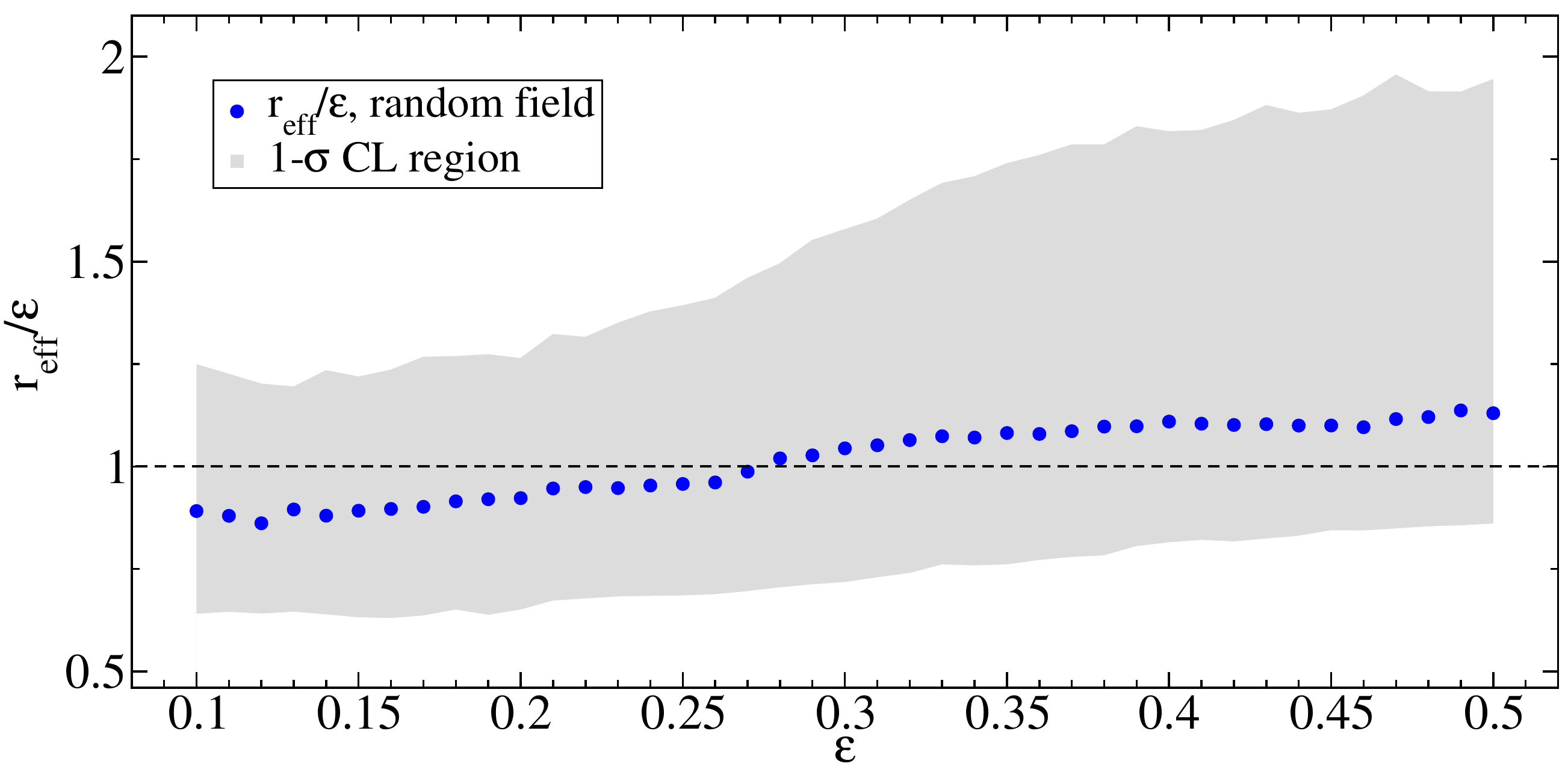}&
\includegraphics[width=9cm]{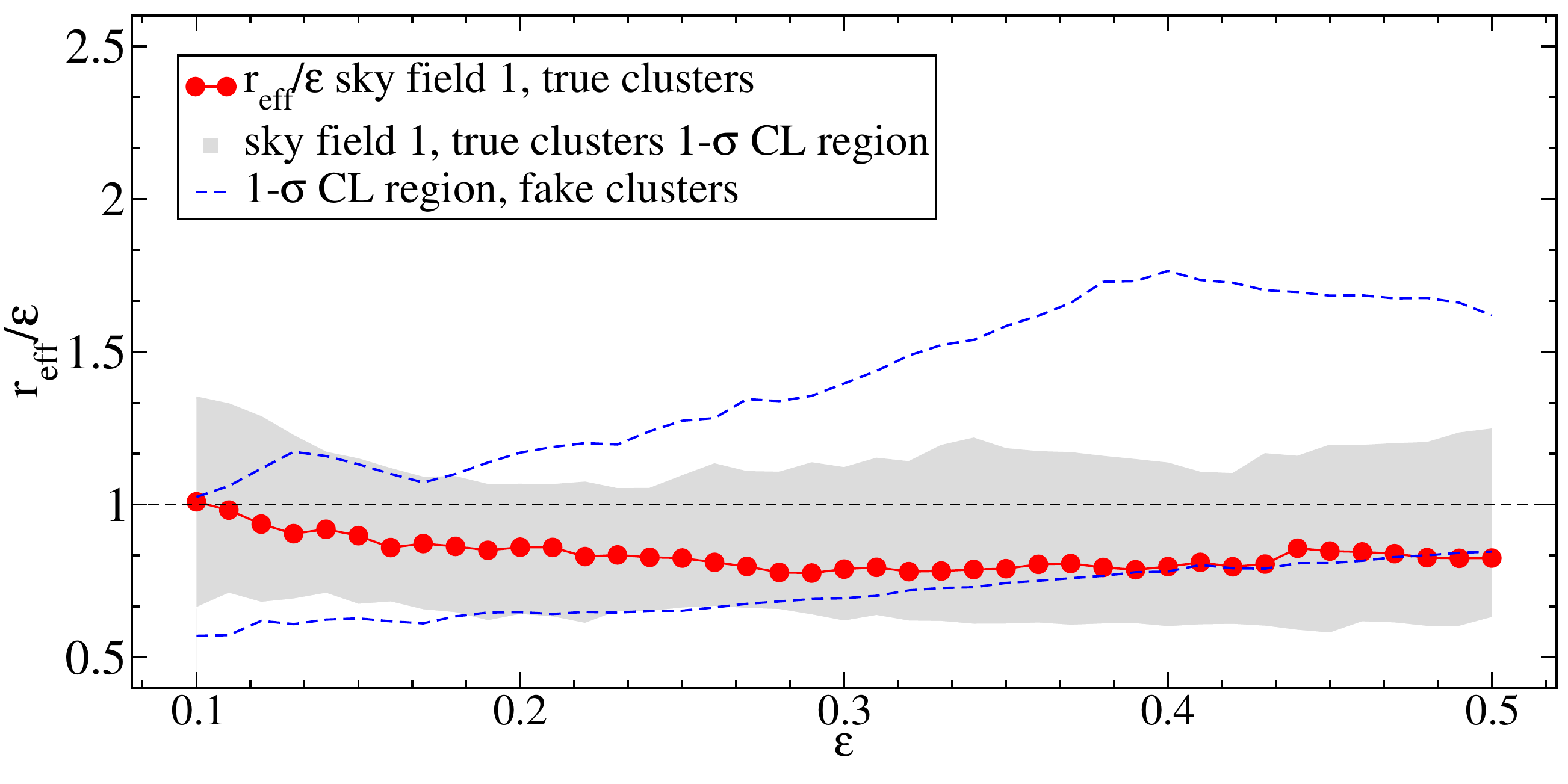}\\
\end{tabular}
\caption{
{\it: Left panel:} the $r_{eff}/\varepsilon$ statistical distribution as a function of \Eps, 
for the {\it random} field case. The blue solid circles represent the median, and the grey shaded 
area represents  the 1-$\sigma$ confidence level region,  for each value of \Eps. 
{\it Right panel:} the same as in the bottom left panel, for the case of the {\it sky} test
field 1. The red solid circles represent the median of the {\it true} clusters case,  and
the grey area the 1-$\sigma$ confidence level region. The  dashed line shows the 
1-$\sigma$ confidence level region, for the case of the {\it fake} clusters.
}
\label{fig:r_68_vs_Eps}
\end{figure*}
	\begin{figure*}
\centering
\begin{tabular}{ll} 
\includegraphics[width=9cm]{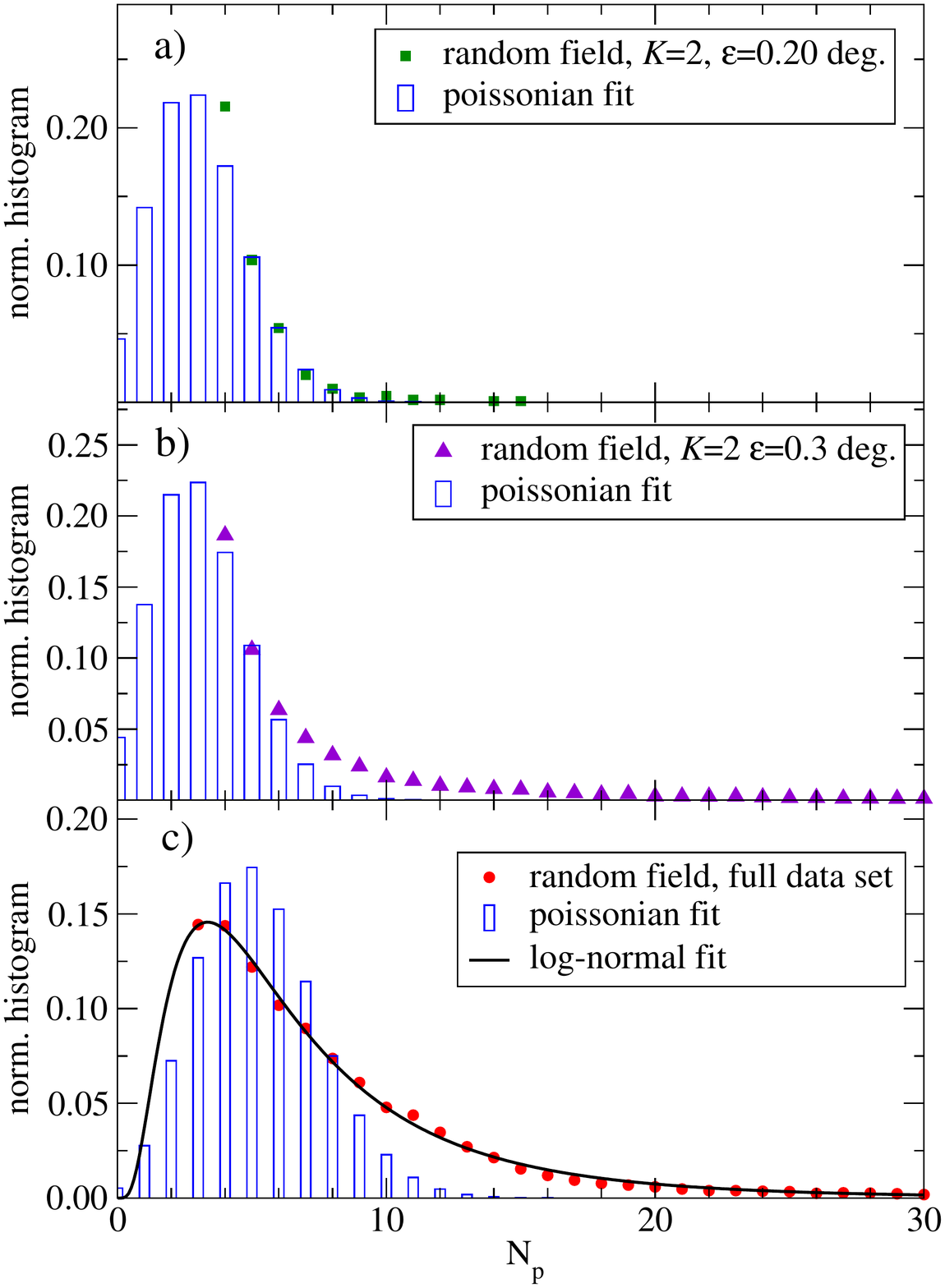}&
\includegraphics[width=9cm]{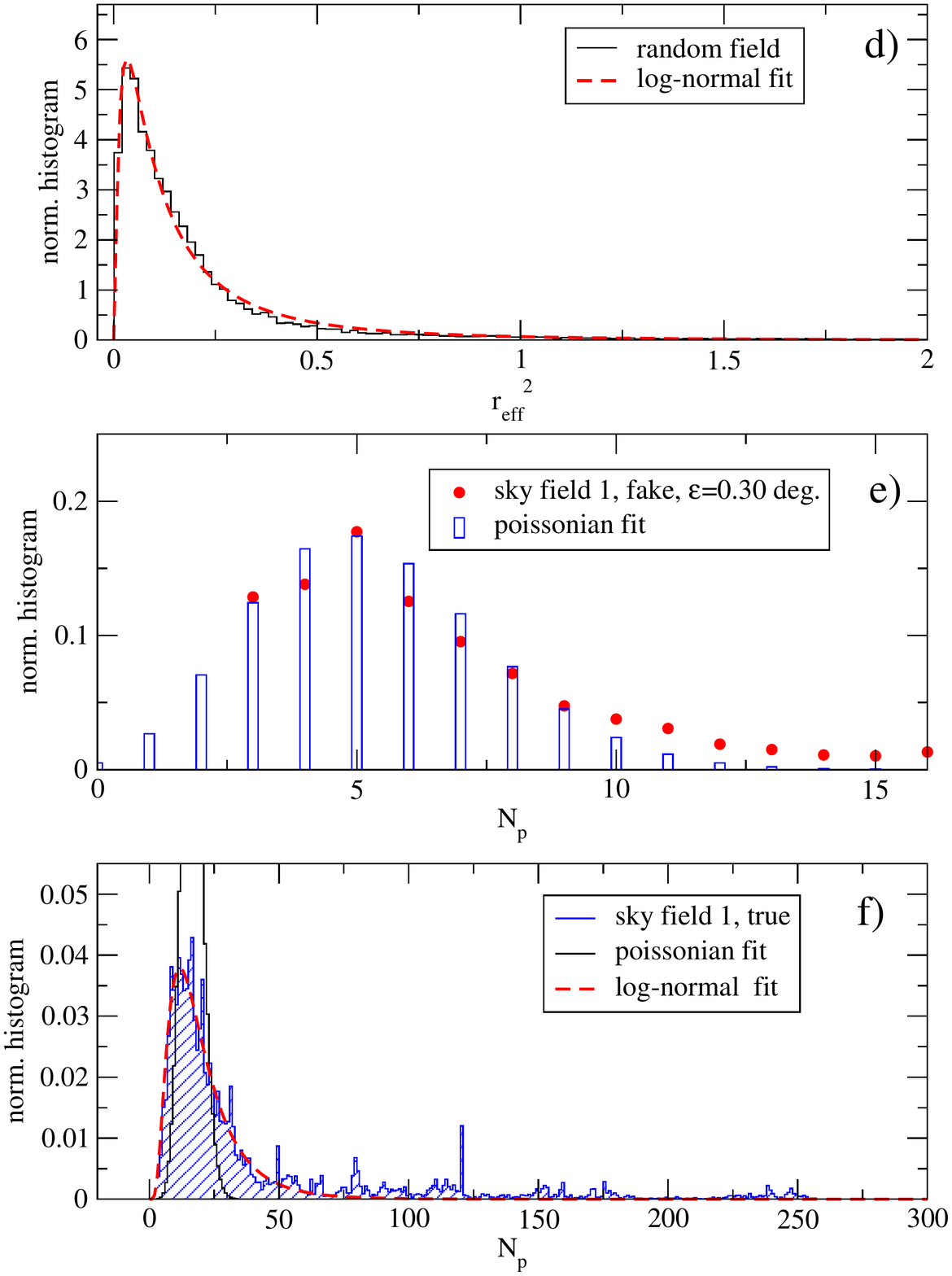}\\
\end{tabular}
\caption{
{\it Left panels:} the distribution of $N_p$ for the {\it random} test field,
for the case of $K=2$, $\varepsilon=0.20$ deg (panel a, red solid boxes). The empty blue bars
line represent a Poissonian best fit. The panel b shows the case of
 $K=2$ $\varepsilon=0.30$ deg (purple solid triangles). 
 The panel c shows the case for the full $K$-\Eps~ parameter space,
 the solid black line represent a  log-normal best fit.
{\it Right panels:} panel c shows the distribution of $r_{eff}^2$ (black solid line),
and it's best fit by means of a log-normal distribution (red dashed line).
Panel  d shows the $N_p$ distribution for the {\it fake} clusters in the {\it sky} test { field} 1 (red 
solid circles),  and the blue empty bars a Poissonian best fit. Panel  e shows the $N_p$ distribution 
for the {\it fake} clusters in the {\it sky} test { field} 1   (blue hatched histogram),  the log-normal best
fit (red dashed line), and the Poissonian fit (solid black line.)
}
\label{fig:Np}
\end{figure*}
\section{Statistical properties of the \gdbscan~ clusters}
\label{sec:stats_sim}
\subsection{The test fields}
In this section we study the statistical properties of the clusters,
looking for signatures that characterize random Poissonian fields,
and fields with point-like sources. To accomplish this task we
compare results obtained for a test field with only noise ({\it random} test
field), and the five test  fields with noise plus point-like sources ({\it sky} test
fields 1-5).

As {\it sky } test fields we use the same fields used in the
\cite{MSTII}. Each of these five {\it sky} fields covers a broad sky region, with
a galactic longitude extension of $80^\circ<l<170^\circ$, and a galactic latitude
extension of $40^\circ<b<65^\circ$. The \gray~ background has been simulated
using the standard {\ttfamily gtobssim}
\footnote{http://fermi.gsfc.nasa.gov/ssc/data/analysis/scitools/help/gtobssim.
txt} tool, developed by the \FLAT~ collaboration, simulating both the Galactic
and isotropic components for a 2-year long period, using a threshold energy of 3
GeV, for a total amount of 9322 photons.  To this photon list  { we added} 70
simulated sources: for each source, the number of photons was chosen from a
probability distribution given by a power-law, with exponent 2, from a minimum
value of 4 up to 40 photons, joined to a constant tail up to 240 photons. 
{ The number of the sources is similar to that reported in  the Fermi-LAT 
Second Source Catalog  \citep[2FGL hereafter]{2FGL}, in the same region of  the sky.} 
The source events are spatially distributed with a bivariate Gaussian 
probability density function (PDF) with $\sigma^{sim}_x=\sigma^{sim}_y = 0.2$ 
deg., centered at the source location. 
Five simulated test fields have been generated, adding the simulated sources to
the diffuse background. The only difference in the five realizations is the
source location,  randomly chosen to have different brightness contrast between
sources and the background.
The {\it random} test field covers the same area of the {\it sky} test fields,
and a number of events equal to  the {\it sky} test field-1 (background and
sources), for a total amount of   11044 events.

In Fig. \ref{fig:sim_detection} we show  the photon map
for the {\it sky} test field 1, and the result of the \gdbscan~ detection
for $K=5$ and $\varepsilon=0.17$ deg. 
We detect 51 {\it true} clusters, and only 2 {\it fake} ones. 
A cluster is defined {\it true}, { if the position of the simulated source  falls within 
a circle centered  on the cluster centroid, with a radius equal to $2pos_{err}$. }
We call {\it fake}, the remaining clusters. 
In Fig. \ref{fig:closeup}, we show a close-up of two
{\it true} clusters. The black ellipses correspond to the ellipses
of the positional error, and the purple  and orange thick points represent the cluster
points, while the black thick dots represent the background.

\begin{figure*}
\centering
\begin{tabular}{ll} 
\includegraphics[width=9cm]{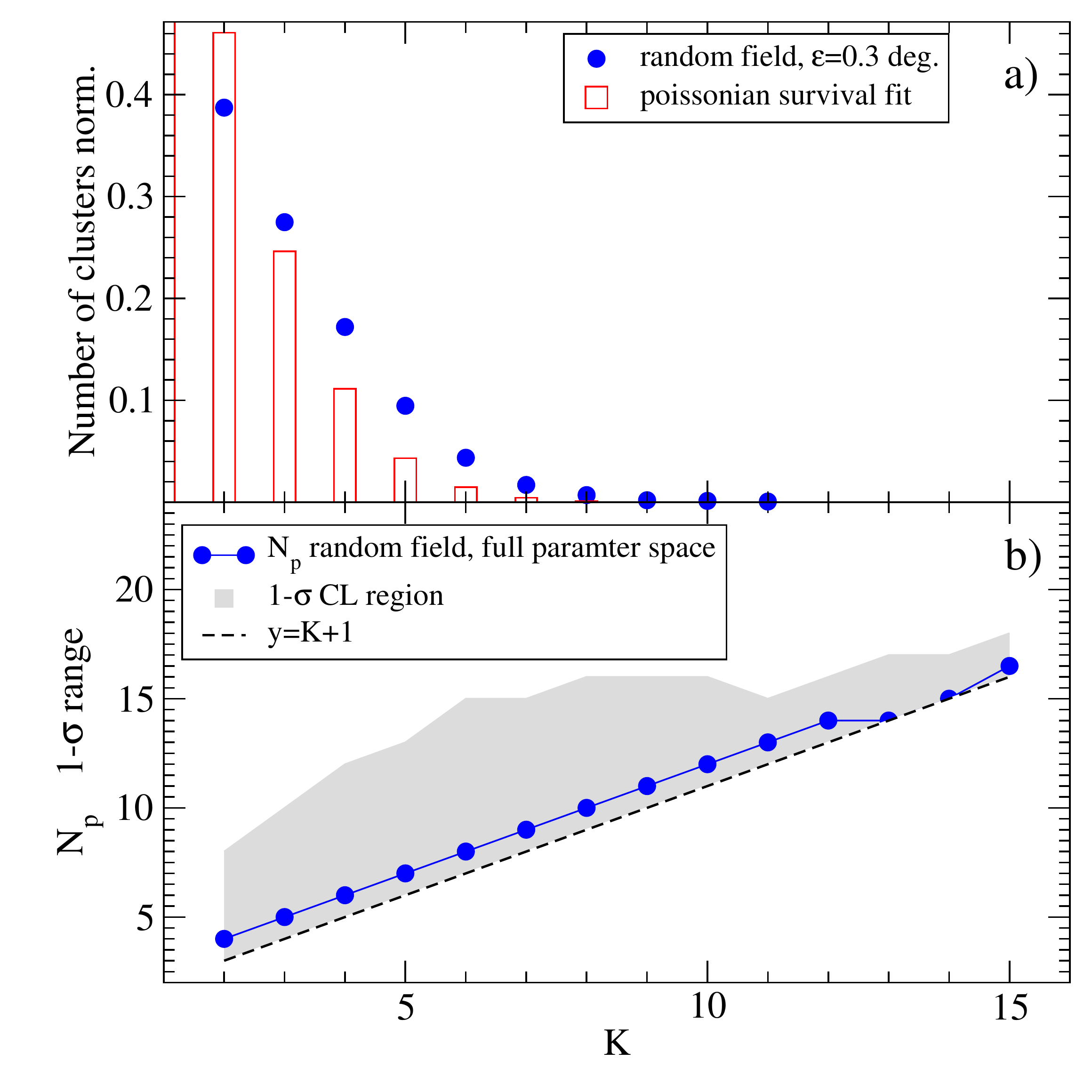}&
\includegraphics[width=9cm]{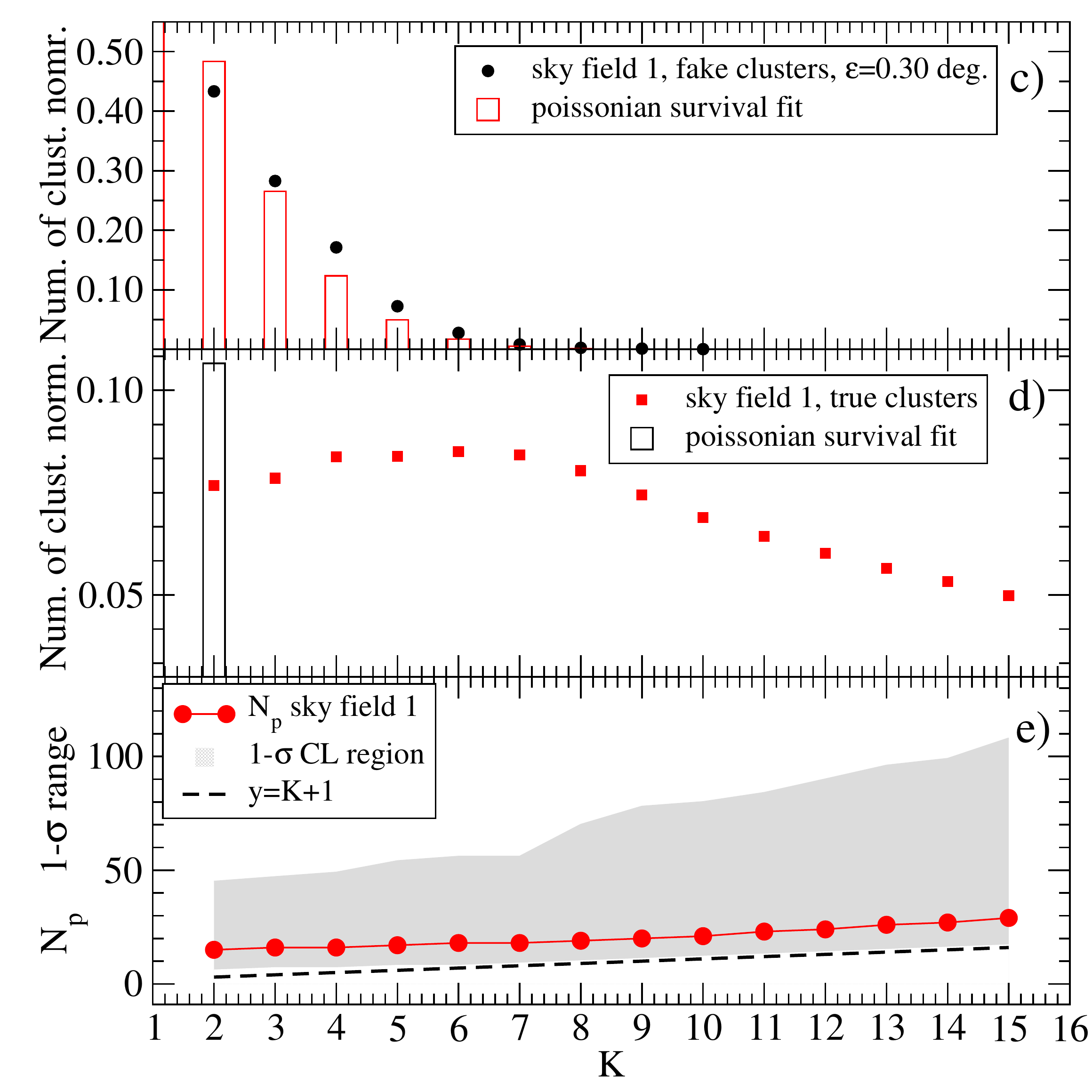}\\
\end{tabular}
\caption{
{\it Panel b:}  the $N_p$ statistical distribution as a function of $K$, for the {\it  
random}  field case. The blue solid circles represent the median, and the grey shaded area represents  
the 1-$\sigma$ confidence level region around the median, for each value of $K$. The dashed black line represents the  
$N_p=K+1$ law.
{\it Panel e:} same as in  panel b, for the {\it sky} test field 1 case.
{\it Panel a:} number of detected clusters for the {\it random} test field case (blue solid   
points), as a function of $K$, and best fit by means of a Poissonian survival  function  
(red empty bars).
{\it Panel c:} number of detected cluster for the {\it sky} test field 1 case (black solid  
points),  for the case of  {\it fake} clusters, as a function of $K$, and best fit by means of a  
Poissonian survival function (red empty bars).
{\it Panel d:} number of detected cluster for the {\it sky} test field 1 case (red solid  
boxes),  for the case of  {\it true} clusters, as a function of $K$, and best fit by means of a  
Poissonian survival  function (black empty boxes).
}
\label{fig:Np_vs_K}
\end{figure*}
\subsection{Test strategy}

We want to investigate the statistical properties of the \gdbscan~
clusters, in particular signatures that distinguish purely random fields
from fields with point-like sources,  and their dependence on $K$ and \Eps. 
To investigate systematically a broad volume of the parameter space, 
we use a parametric approach. We set the range of \Eps~ in [0.1$\div$0.50] deg. with a 
step of 0.01 deg., and the range of $K$ in [2$\div$15],  with a step of 1. The
total amount of detection trials for each test field is 574. 
We collect the statistics of the trials, and we investigate the distribution
of  $r_{eff}$ and $N_p$, and their connection with \Eps~ and $K$, respectively.

\subsection{Statistics of $r_{eff}$ and connection with \Eps}
We start by investigating the distribution of the $\log(r_{eff})$ values, in the
case of the  {\it random}, and of the {\it sky} test field-1. The distribution for the
detections collected over the full $K$-\Eps~  parameter space (top left panel of
Fig. \ref{fig:r_68}),  shows a symmetric shape 
well fitted by a Gaussian distribution (log-normal w.r.t. $r_{eff}$), with
the mean value of $<\log_{10}(r_{eff})>\simeq -0.45$ (corresponding to
$<r_{eff}>\simeq 0.3$ deg) and a dispersion of $\sigma_{\log_{10}( r_{eff})}\simeq$ 0.23.
{
The log-normal distribution  provides a reasonable
description of the empirical distributions also for individual pairs of ($K$,\Eps) values.
An example is
given in panel c of  Fig. \ref{fig:r_68}, for the case $K=3$, $\varepsilon=0.3$ deg.,
where the best fit values are $<\log_{10}(r_{eff})>\simeq -0.51$, and $\sigma_{\log_{10}( r_{eff})}\simeq$ 0.16.
We now investigate the empirical distribution of  $\log_{10}(r_{eff})$ for 
fields with point-like sources. In the right panel of Fig. \ref{fig:r_68},
we show the case of  the {\it  sky}  test field-1. The distributions of $\log_{10}(r_{eff})$ are 
still described by a by a normal. In the case of {\it fake} clusters (red dashed line), the 
best fit values of the mean ($<\log_{10}(r_{eff})>\simeq -0.46$) and of the
dispersion ($\sigma_{\log_{10} r_{eff}}\simeq$ 0.24), are very similar to those found  in the
case of the {\it random} test field. On the contrary, the {\it true} cluster distribution (blue hatched histogram) 
is peaking around the value of $\log_{10}(r_{eff})\simeq-  0.67$ deg, corresponding 
to $r_{eff}\simeq 0.21$. deg., very close to  the value of the 
dispersion $\sigma^{sim}=0.20$ deg., used to simulate the sources. }
Since the simulation parameter $\sigma^{sim}$ reproduces the
effect of the instrumental PSF, we observe that for non-{\it random} fields, the
typical size of the reconstructed clusters is constrained by the PSF, suggesting
the empirical rule to set the value of \Eps~ of the order of the PSF size.

To investigate more accurately the connection between \Eps~ and the PSF, we analyse 
the statistical properties of the quantity  $r_{eff}/\varepsilon$ as a function of \Eps. 
For each value of \Eps, we determine the median, and the  two-sided 1-$\sigma$ 
confidence level (CL) interval around the median,  of the $r_{eff}/\varepsilon$ distributions.
In the left  panel of Fig. \ref{fig:r_68_vs_Eps} we plot the $r_{eff}/\varepsilon$ 
median   (blue solid circles) and 1-$\sigma$  CL region, as  a function of $\varepsilon$,
for the {\it random} field.  
We note that  the $r_{eff}/\varepsilon$ trend is slightly increasing with \Eps, 
and that the 1-$\sigma$ CL region is consistent with the case $r_{eff}/\varepsilon=1$,
but  the upper boundary shows a systematic increase, compared to the lower
boundary, for \Eps$\gtrsim 0.30$ deg.
{ The trend for the case of the {\it true} clusters in {\it sky} test field 1 
(right panel Fig.\ref{fig:r_68_vs_Eps}), shows a different behaviour. 
The median of $r_{eff}/\varepsilon$ (red solid circles) is slightly decreasing with
\Eps, showing that, for {\it true } clusters, $r_{eff}$ is not sensitive to the 
size of \Eps, being mostly constrained by the simulated PSF size. As expected,  for the case of 
{\it fake} clusters (blue dashed line), the trend is almost identical to that 
of the clusters in the {\it random} field.
}

\subsection{Statistics of $N_p$ and connection with $K$}
We now investigate the statistics of the distribution of the number of 
photons per cluster. In the case of {\it random} fields, we expect  that
the number of photons in a cluster attends a Poisson distribution. Indeed,
for a generic two-dimensional Poisson process,  the probability to observe
a number of events ($N(S)=j$) enclosed by a surface $S$  is given by:
\begin{equation}
P(N(S)=j)=\frac{(\lambda |S|)^j \exp{(-\lambda |S|)}}{j!},
\label{eq:Poisson}
\end{equation}
where $\lambda$ is the average spatial density.
Translating $S$ in terms of $\varepsilon^2$, we can rewrite:
\begin{equation}
P(N(\varepsilon^2)=j)=\frac{(\lambda |\varepsilon^2|)^j \exp{(-\lambda |\varepsilon^2|)}}{j!},
\label{eq:Poisson_Eps}
\end{equation}
from which follows that, given the value of $K$ and $\varepsilon$, the probability
to find a cluster  as function of $K$ and \Eps~ will be given by
\begin{equation}
P_{clus}(\varepsilon,K)=P(N(\varepsilon^2)>K)=1- \sum\limits_{j=0}^{K} \frac{(\lambda |\varepsilon^2|)^j\exp{(-\lambda |\varepsilon^2|)}}{j!},
\label{eq:P_clus}
\end{equation}
namely the Poissonian  survival function. 
Anyhow, due to  the logic of the  \dbscan~ 
clustering process, the Poisson statistics can't be extended  from \Eps~ 
to $r_{eff}$, for any value of \Eps.
Indeed, a cluster  is not a simple collection of points enclosed
within a surface $S$, this holds only within the \Eps-sized circle, namely the
{\it seed} of the cluster ($C^*$). If we consider the annulus defined  between \Eps~ 
and the cluster radius  $r_{clus}$, not all the points in the annulus will be cluster
member, but only those that are at least density reachable. 
This implies that we expect a deviation from the Poisson 
statistics, when $r_{eff}$  is significantly larger than \Eps, i.e. \Eps $\gtrsim$ 0.3 deg. 
(according to the analysis presented in the previous section).  This expected deviation from the 
Poissonian statistics,  is confirmed by the plots in the left panels of Fig. \ref{fig:Np}.
In panel a we show the distribution of $N_p$ for the
case $K=2$ and $\varepsilon=0.20$ deg. We note that the Poisson distribution (Eq. \ref{eq:Poisson_Eps}) 
gives a reasonable  description of the empirical distribution. On the contrary, for the case of
$\varepsilon=0.30$ deg. (panel b), we observe that the Poisson distribution shows larger deviations,
in particular for $K>6$.  When we take into account the $N_p$ distribution for the full 
parameter space (panel c), we note the Possonian distribution is failing in  providing
a reasonable description of the empirical distribution, whilst a log-normal one gives a 
good fit.

\begin{figure*}
\centering
\includegraphics[width=18cm]{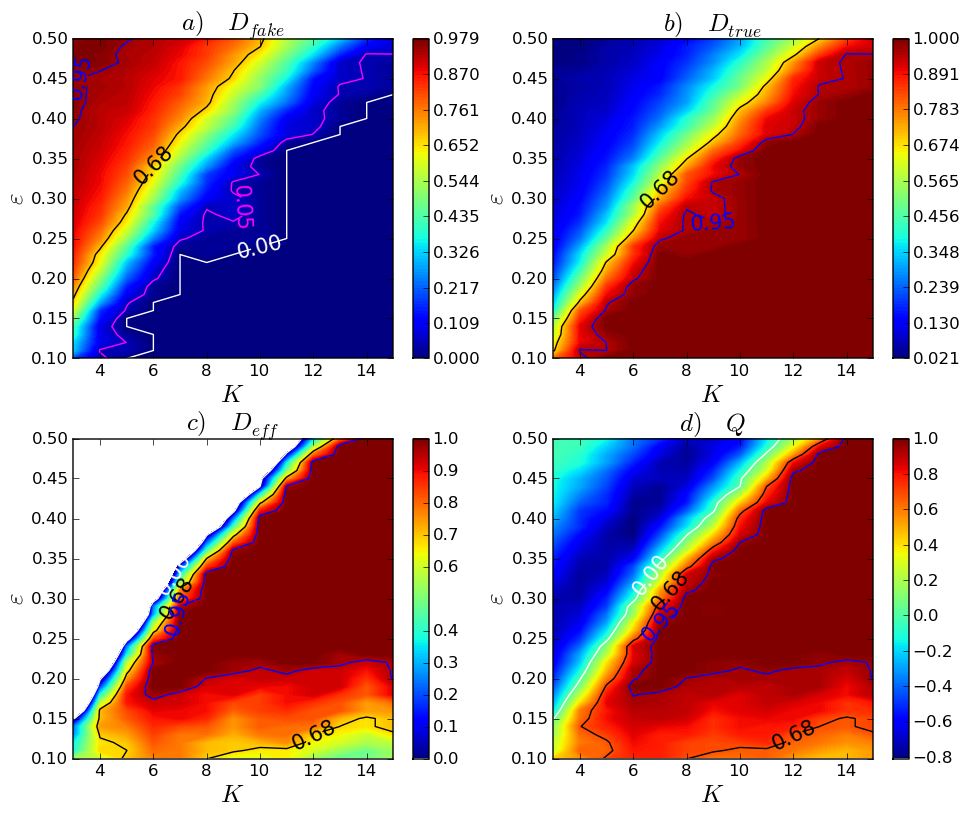}
\caption{
Isolevel maps for $D_{fake}$ (panel a),  $D_{true}$ (panel b), $D_{eff}$ (panel c), and $Q$ 
(panel d), for the {\it sky} test field 1. The white lines show the
$\mathrm{isolevel}=0$, the black lines show the $\mathrm{isolevel}=0.68$, and the blue lines 
show the $\mathrm{isolevel}=0.95$.
}
\label{fig:param_run}
\end{figure*}

The log-normal trend of $N_p$ is consistent with the log-normal trend of the  
distribution of $r_{eff}$.  Since the number of photons in a cluster will be 
approximatively $N_p \propto \lambda r_{eff}^2$,
we can write the PDF of $N_p$:

\begin{equation}
f(N_p)\propto f(r_{eff}^2)\lambda.
\label{eq:f_N_p}
\end{equation}

To evaluate the distribution of $ r_{eff}^2$ we can use the
standard theory of the transformation of Random Variables (RV) \citep{papoulis}.
It can be easily proved that, given a RV $X$ having a log-normal distribution,

\begin{equation}
f(X)=\frac{1}{X\sqrt{2\pi\sigma^2}}\exp\Big({\frac{-\ln(X)-\mu}{2\sigma^2}}\Big), 
\end{equation}
the RV $Y=X^2$, will follow a log-normal distribution given by:

\begin{equation}
f(Y)=\frac{1}{2Y\sqrt{2\pi\sigma^2}}\exp\Big({\frac{-\ln(Y)-2\mu}{4\sigma^2}}\Big). 
\end{equation}			

Indeed, our $r_{eff}^2$ distribution, for the {\it random} field (panel d, 
Fig. \ref{fig:Np}), is fitted by a log-normal distribution peaking at $\simeq 0.03$ deg$^2$.
Hence, according to Eq. \ref{eq:f_N_p} we expect that also $f(N_{p})$ will follow a log-normal
distribution, when $N_p$ is not ruled by a Poissonian statistics.

We verify, that the same statistical trends, describe the {\it real} sky fields. 
The  panels e and f in Fig. \ref{fig:Np}, show 
the statistical distribution of $N_p$ for the {\it sky} test field 1 case. 
In agreement with the analysis concerning the {\it random} test field, 
we see that the {\it fake } clusters  ($\varepsilon=0.30$ deg., panel e in 
Fig. \ref{fig:Np}), are described  by a Possonian statistic, whilst, the  {\it true } 
clusters (panel f in Fig.  \ref{fig:Np}), are  better described by a log-normal 
distribution (red dashed line), compared to a Poissonian  one (solid black line). 
We  also observe that the log-normal law,  describes reasonably 
the empirical distribution, only for values of $N_p\lesssim 50$, whilst shows 
significant deviation  in the tail, consistent with the statistics of our simulated sources    
population.

To complete this statistical characterization, we investigate the distribution
of the number of detected clusters, as a function of the threshold $K$.
According to Eq. \ref{eq:P_clus}, we expect that the number of detected cluster,
for a {\it random} field, follows a Poisson survival distribution. The  plot a of 
Fig. \ref{fig:Np_vs_K} confirms our hypothesis, indeed
the Poisson  survival function provides a reasonable description of the 
empirical distribution. The same holds for the case of {\it fake} clusters of
the {\it sky} test field 1 (plot c Fig. \ref{fig:Np_vs_K}). On the
contrary, in the case of {\it true} clusters (panel d Fig.
\ref{fig:Np_vs_K}), the Poisson survival distribution is not able to reproduce
the observed trend, consistently with the non-poissonian statistic of the
simulated clusters . The panels b and e of Fig. \ref{fig:Np_vs_K}, show the
$1-\sigma$ CL region for the $N_p$, as a function of $K$. We note, that both in the
case of {\it random} and {\it sky} field {\it true} clusters, the lower boundary
of the region is constrained by the equation $y=K+1$, that is consistent with
the \gdbscan~ logic.
On the contrary, the upper boundary shows a different
behaviour. In the case of the {\it random} field, the upper boundary
deviates from the lower boundary compatibly with the fluctuations 
of the events around the \Eps~ circle, and ranges from about 8 to about 16.
On the contrary, in the case of {\it sky} field {\it true} the upper boundary
is constrained by the statistics of the number of events in the simulated sources,
and rages from about 60 to 100.

\section{Testing  the detection performance with simulated \gray~ data}
\label{sec:det_sim}
In this section we investigate the detection performance of the
\gdbscan.
As first point, we study the dependency of the detection efficiency on $K$ and \Eps, 
and their impact on the spurious ratio, and on the detection efficiency. 
Then, we investigate the capability of
the algorithm to reconstruct the simulated clusters, and the positional accuracy
of the reconstructed centroids.  
We test the detection performance of the \gdbscan~ using as benchmark the 
five {\it sky} test fields used in the previous section, and exploring
the same parameter space.

\subsection{Detection efficiency and spurious ratio as a function of $K$ and 
\Eps}

To investigate the detection performance of the \gdbscan,  we   run, 
for each of the five {\it sky} test fields, and  for each pair of values $K$,\Eps,
a \gdbscan~ detection. For each detection run, we build a  {\it cluster catalog}.
Starting from the {\it cluster catalog}, we build the corresponding {\it candidate catalog}. 
The {\it candidate catalog} is a list of  sources built by  taking into  
account two possible biases, the {\it confusion}, and the {\it multiple association}, 
in detail:
\begin{itemize}
\item{} a cluster is defined {\it true}, i.e. with a possible counterpart, 
{    if the position of the simulated source 
 falls within a circle centered  on the cluster centroid, with a radius equal to 
 $2 pos_{err}$. }
\item{} Two or more {\it true} clusters are defined {\it confused}, if they have the same counterpart
\item{} A {\it true} cluster has a  {\it multiple association}, if has more than one counterpart.
\end{itemize}
We stress that, the number of confused clusters is negligible, 
{ indeed the average number of 
{\it confused} clusters per run is about 0.08, and no  {\it confused} clusters 
are found for $K>4$, and that  the average number of  {\it multiple associations}
per run is about 0.2.
}

The final {\it candidate catalog} will count a number  of candidate sources $N_{src}$,  
each identified by a unique $SRC_{ID}$.  The  number of spurious sources  will be 
$N_{fake}=N_{src}-N_{true}$. In order to to characterize the performance, we
define the following parameters:

\begin{itemize}
\item{} the detection efficiency:
\begin{equation}
D_{eff}= 
\left\{
\begin{array}{c l}
\frac{N_{true}-N_{fake}}{N_{sim}(N_p sim.>K)}, &\mathrm{if~}  (N_{true}-N_{fake}) \leq N_{sim}(N_p sim.>K) \\
1.0                                          , &\mathrm{if~}  (N_{true}-N_{fake})>N_{sim}(N_p sim.>K) \end{array} 
\right.
\label{eq:Deff}
\end{equation}
where $N_{sim}(N_p sim.>K)$ is the number of simulated sources
with a number of simulated events larger than $K$
\item{} the true detection ratio  $D_{true}=N_{true}/N_{src}$
\item{} the spurious detection ratio  $D_{fake}=N_{fake}/N_{src}$
\item{} the overall detection quality factor ($Q$), that takes into account 
the tradeoff between $D_{eff}$  and $D_{fake}$, defined as:
\begin{equation}
Q=D_{eff}\Big(1-\frac{N_{fake}}{N_{src}}\Big)
\label{eq:Qeff}
\end{equation}
\end{itemize}

\begin{figure*}
\centering
\begin{tabular}{ll} 
\includegraphics[width=9cm]{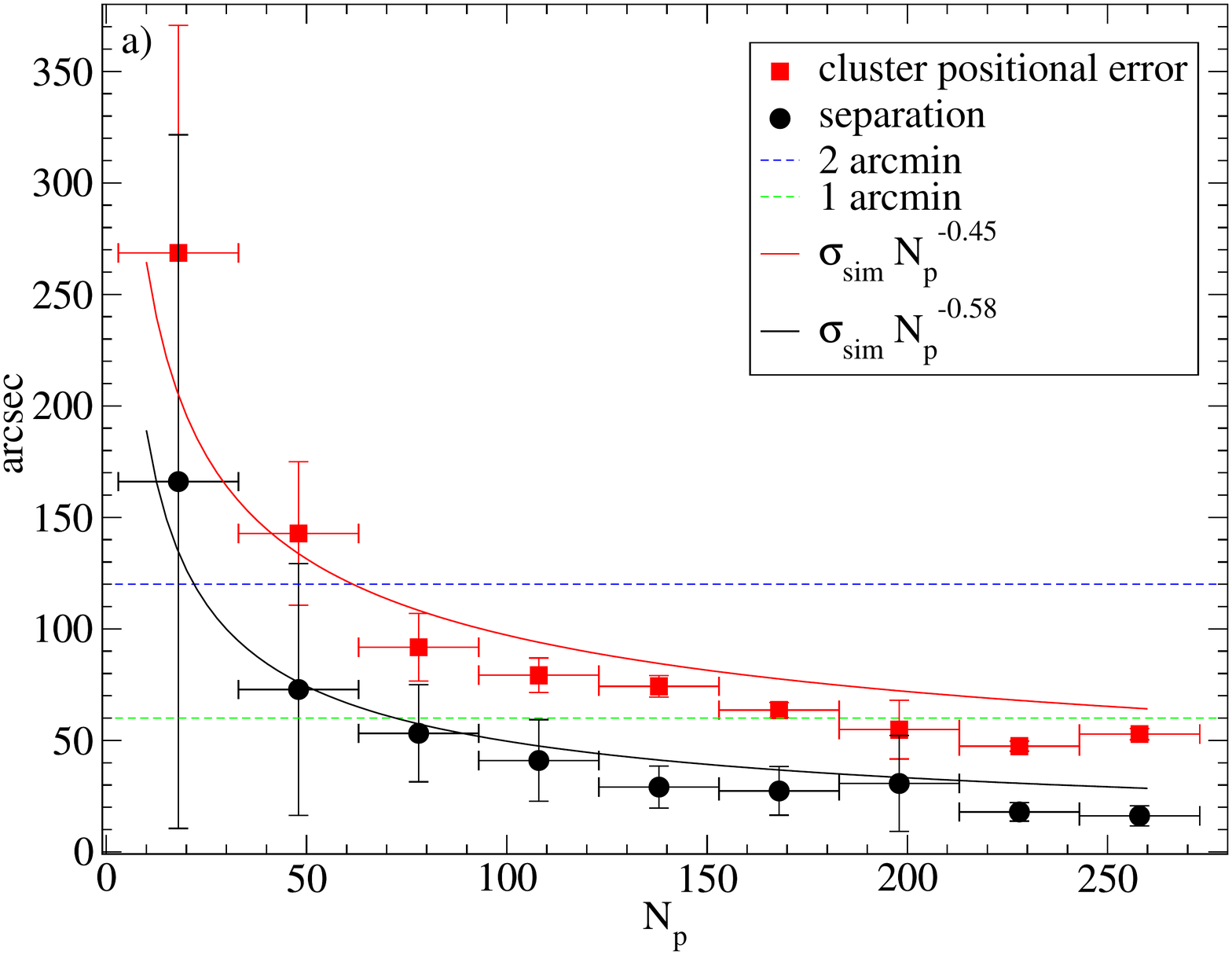}&
\includegraphics[width=9cm]{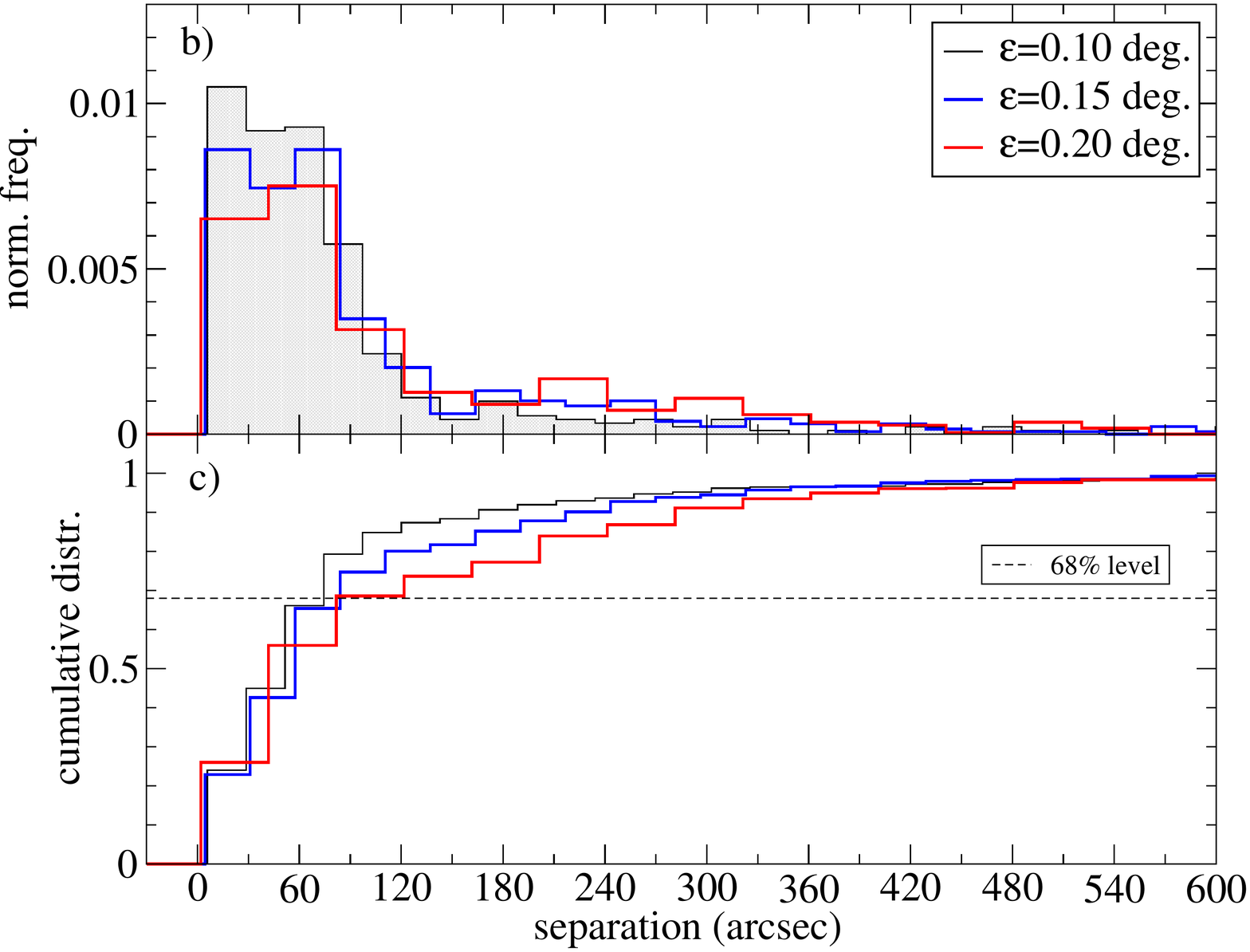}\\
\end{tabular}
\caption{
{
{\it Panel a:} red solid boxes show the mean positional error of the centroid, for
{\it true} clusters in {\it sky} test field 1, and the standard deviation (vertical error bar), vs. $N_p$.  
The clusters are  binned in $N_p$, with the bin width indicated by the horizontal error bar. 
The black solid circles represent the corresponding trend for the distance between the cluster 
centroid and the simulated source position.
} 
{\it Panel b:} the distribution of the distance between the simulated source position and the 
cluster centroid, expressed in arcsec, for the case of $\varepsilon=0.10$ deg. (black line), 
$\varepsilon=0.15$ deg. (blue line), and $\varepsilon=0.20$ deg. (red lines). {\it Panel c:} the cumulative 
distributions  corresponding to panel b.
}
\label{fig:Pos_prec}
\end{figure*}
\begin{figure}
\centering
\includegraphics[width=9cm]{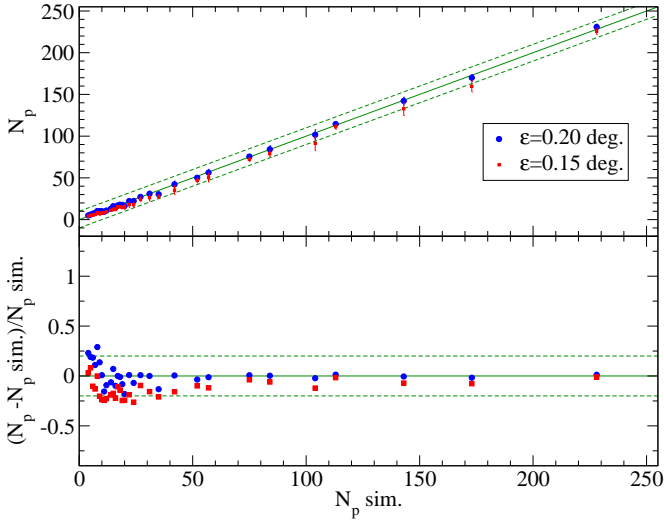}
\caption{
{
{\it Top panel:} the average number  of photons associated to each clusters 
$N_p$, and their dispersion (vertical bar) vs. the number of photons  
simulated  ($N_p$ sim). The red points refer to
the sub parameter space $\varepsilon=0.15$ deg., and the solid blue  circles to the
$\varepsilon=0.20$ deg. sub space. The solid green lines represent the
law  $N_p=N_p$ sim. The dashed lines represent the law $N_p=N_p$ sim. $\pm$ 10.
{\it Bottom panel:} The corresponding fractional deviation $(N_p-N_p sim.)/N_p sim$.
}
}
\label{fig:Np_rec}
\end{figure}

{ The $D_{eff}$ parameter shows the fraction of simulated clusters, 
above the threshold $N_p sim=K$, detected by the method,
net of the {\it fake} ones. Hence, does not provide an indication
of the spurious contamination. For this reason we have introduced
the $Q$ parameter, which rescale the $D_{eff}$ according to the ratio 
between {\it fake} clusters, and found clusters $N_{src}$.}
We remind that, according to the $D_{eff}$ definition in Eq. \ref{eq:Deff},
it's possible to obtain values of $D_{eff}>1.0$. Assume to have a simulated
cluster such that, for a given $K$ and \Eps,  the corresponding seed cluster  
has a size $N^*=N_p sim.=K$. In the case of no background events within
the circle of radius \Eps, this cluster  will be rejected. 
If we have one or more background events contained within the circle	
of radius \Eps, i.e.  $N^*>K$, the cluster will be detected.
For this reason, in such a case, we report a value of $D_{eff}=1.0$. 
The same applies to $Q$.

In Fig.  \ref{fig:param_run} we summarize the detection runs for the 
case of {\it sky} test field 1, for the full parameters space with 
$K>2$. The  panel a shows the isolevel map of the {\it fake} 
clusters detection ratio. The gradient in the isolevel map is quite 
sharp, and  roughly half of the parameter space shows no {\it 
fake} clusters (white isolevel line). To have a better understanding of 
the impact of {\it fake} clusters, it's interesting to compare the 
$D_{fake}$ isolevel map  to the $D_{true}$ isolevel map (panel b 
Fig. \ref{fig:param_run}). Also in this case the map shows a 
sharp  gradient, and the region with $D_{true}>0.95$ overlaps the 
$D_{fake}=0$ region. These two maps, clearly show the region
of the parameter space where the algorithm has the best performance,
but the $D_{true}$ and $D_{fake}$ ratios do not provide information on the { ratio} 
between the number of {\it true} detected clusters and the number
of simulated clusters. At this regard more information are provided
by the  $D_{eff}$ isolevel map (panel c, Fig. \ref{fig:param_run}). To focus on the 
"effective" volume of the parameter space,  we hide by a white area 
the region where $D_{eff}<0$. 
We note that the isolevel lines $D_{eff}=0$ 
and the isomap lines in the maximum gradient area show a positive 
correlation between  $K$ and \Eps, meaning that  an increased  value of \Eps~, 
requires an increased value of $K$, to have better background rejection. 
To evaluate better the trade-off 
between $D_{true}$ and $D_{fake}$, we plot in the panel d  
of Fig. \ref{fig:param_run}, the isolevel map of $Q$. This plot 
shows that the area corresponding to $Q>0.95$, is consistent with that found in the 
case of $D_{eff}$.
In  Tab. 1 we report the $D_{eff}$ values obtained for all the five 
{\it  sky} fields, for  detections with a number  of {\it fake}  sources 
$\leq$ 6.  We note that the average values of {\it true} clusters ranges
between 44 and 51, with the {\it fake} ones ranging between 1 and 3, 
and an average  $D_{eff}$ between 0.96 and 1.0.
This is a very promising result.  

\subsection{Cluster reconstruction, and positional accuracy}
The positional accuracy of the topometric methods, is probably
the most important feature of this class of algorithms.
In Sec. \ref{sec:dbscan}, we have described our weighting 
method to reconstruct the centroid of the cluster.

{
The panel a of Fig. \ref{fig:Pos_prec} shows by red solid boxes
the mean positional error of the clusters centroid and the standard 
deviation (vertical error bar) vs. $N_p$, for the {\it true} clusters 
of the {\it sky} test field 1 with  $\varepsilon \leq 30$ deg.
The clusters are binned in $N_p$, with the bin width indicated by the horizontal 
error bar. 	
As expected, the uncertainty on the reconstructed cluster centroid 
is  $pos_{err} \approx \sigma_{sim}/\sqrt N_p$ (solid red line).
The solid black circles represent the corresponding trend 
for the separation between the simulated cluster position
and the reconstructed cluster centroid. For $N_p \gtrsim 30$,
the separation is below $2'$. 
In  the panel b   of Fig. \ref{fig:Pos_prec} we plot the
histogram of the distribution of the angular separation 
between the position of the simulated source, and the
position of the cluster centroid. 
For the three cases of $\varepsilon=0.10$ deg., $\varepsilon=0.15$ deg.,  
and  $\varepsilon=0.20$ deg., the positional error is below the $1.5'$, for 
the $68\%$ of the sample. 
}

\begin{table}

\begin{tabular}{@{}r|ccccccc@{}}
\hline
  &$N_{sim}^{cut}>K$ & True & Fake & $K$ & \Eps~ & $D_{eff}$ & $Q$  \\ 
\hline
\hline 
    
{\it sky} field 1 &44/70  &47 &1 &6  &0.19  &1.00 &1.00 \\ 
                   &38/70  &48 &2 &8  &0.27  &1.00 &1.00 \\ 
                   &53/70  &50 &3 &5  &0.17  &0.89 &0.84 \\ 
                   &62/70  &51 &5 &4  &0.14  &0.74 &0.68 \\ 
                   &53/70  &53 &6 &5  &0.18  &0.89 &0.80 \\ 
\hline
average   &50.00 &49.80 &3.40 &5.60 &0.19 &0.90 &0.86 \\ 
\hline
\hline
{\it sky} field 2 &62/70  &47 &0 &4  &0.11  &0.76 &0.76 \\ 
                   &44/70  &49 &2 &6  &0.20  &1.00 &1.00 \\ 
                   &53/70  &51 &3 &5  &0.17  &0.91 &0.86 \\ 
                   &62/70  &52 &6 &4  &0.14  &0.74 &0.67 \\ 
\hline
average   &55.25 &49.75 &2.75 &4.75 &0.15 &0.85 &0.82 \\ 
\hline
\hline
{\it sky} field 3 &41/70  &42 &0 &7  &0.22  &1.00 &1.00 \\ 
                   &44/70  &45 &2 &6  &0.20  &0.98 &0.94 \\ 
                   &53/70  &46 &3 &5  &0.16  &0.81 &0.76 \\ 
                   &62/70  &48 &5 &4  &0.14  &0.69 &0.63 \\ 
                   &62/70  &50 &6 &4  &0.15  &0.71 &0.63 \\ 
\hline
average   &52.40 &46.20 &3.20 &5.20 &0.17 &0.84 &0.79 \\ 
\hline
\hline
{\it sky} field 4 &53/70  &47 &1 &5  &0.16  &0.87 &0.85 \\ 
                   &53/70  &50 &5 &5  &0.18  &0.85 &0.77 \\ 
                   &62/70  &52 &6 &4  &0.14  &0.74 &0.67 \\ 
\hline
average   &56.00 &49.67 &4.00 &4.67 &0.16 &0.82 &0.76 \\ 
\hline
\hline
{\it sky} field 5 &44/70  &47 &1 &6  &0.19  &1.00 &1.00 \\ 
                   &44/70  &50 &2 &6  &0.20  &1.00 &1.00 \\ 
                   &44/70  &53 &3 &6  &0.21  &1.00 &1.00 \\ 
                   &44/70  &55 &5 &6  &0.22  &1.00 &1.00 \\ 
\hline
average   &44.00 &51.25 &2.75 &6.00 &0.20 &1.00 &1.00 \\ 
\hline
\hline

\hline  
\end{tabular}

\label{tab:tab1}
\caption{
Summary of the	detections obtained for all the five {\it sky} fields, 
for detections with a number of fake sources $\leq 6$. $N_{sim}^{cut}>K$, is is the number of 
simulated sources with a number of simulated events larger than $K$, the number separated by the 
$/$ symbol, indicates the full number of simulated sources.
}
\end{table}

Besides positional accuracy, is also important to understand
the capability of the \gdbscan~ to reconstruct the simulated
cluster in terms of number of photons. Indeed, this information
gives an idea of the average number of background photons
contaminating the reconstructed cluster. In the top left panel 
of Fig. \ref{fig:Np_rec}, we show the scatter plot
of $N_p$ vs. the number of simulated events ($N_p$ sim.). 
The solid points represent the average value of $N_p$,
for a given value of  $N_p$ sim., and the error bar, corresponds 
to the standard deviation.
The solid green line represents the case $N_p=N_p$ sim.,
and the dashed upper and lower lines, represent $N_p=N_p\pm 10$ sim.,
respectively. 
Both for the cases of $\varepsilon=0.15$ deg.,  and  $\varepsilon=0.20$ deg.,
the scatter is bounded by the dashed lines, showing that the largest
excess in the $N_p$ is about 10 photons, independently of $N_p$ sim.
We note, that in the case of $\varepsilon=0.15$ deg., the number of reconstructed
photons, systematically underestimates the simulated number, whilst, 
the $\varepsilon=0.20$ deg. case does not shows this bias.
It's possible to appreciate better this effect, in the  bottom left panel   
of Fig. \ref{fig:Np_rec}, where we show
the fractional reconstruction error ($N_p-N_p$ sim.)/$N_p$ sim., vs. 
$N_p$ sim. The solid green line represent the case with 0 error, 
and the dashed lines represent the $\pm 20\%$ boundaries. 
The bias on $N_p$ in the case of $\varepsilon=0.15$ deg., shows again
the strong correlation between \Eps~ and the PSF radius. When \Eps~ is
smaller then the $\sigma^{sim}$ (that in our simulations reproduces
the PSF effect), the number of reconstructed events $N_p$ is systematically
smaller than $N_p$ sim., on the contrary, when the \Eps~ radius matches the PSF
radius size ($\varepsilon=0.20$ deg.), the bias disappears.

\begin{figure*}
\centering
\begin{tabular}{ll} 
\includegraphics[width=9.0cm]{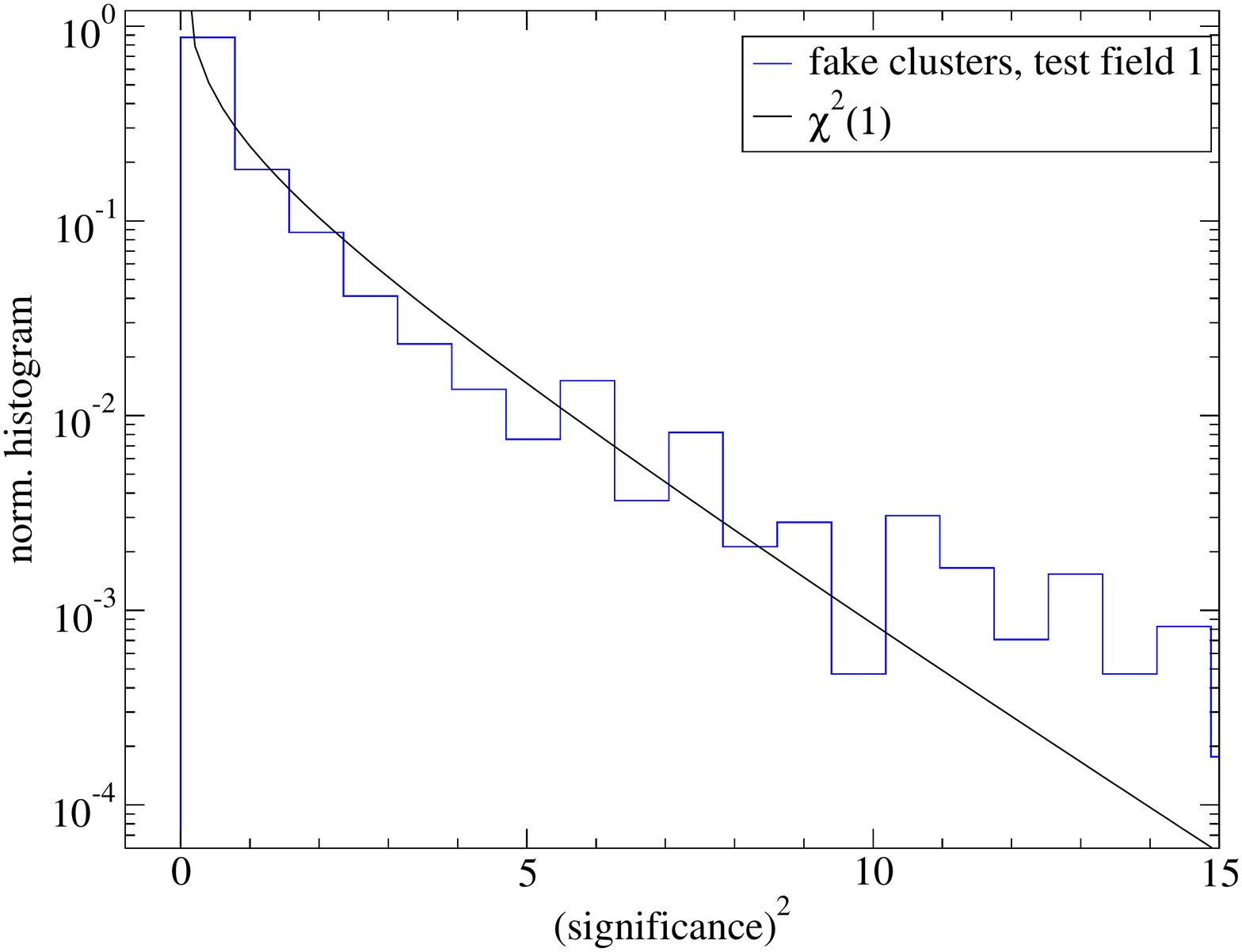}&
\includegraphics[width=8.5cm]{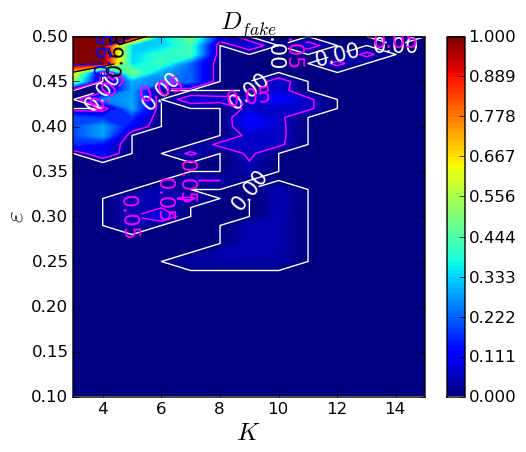}\\
\end{tabular}
\caption{
{\it Left panel:} The distribution (blue line) of the square of the significance, for 
the {\it fake} clusters in the {\it sky} test field 1, for the full $K$,\Eps~ parameter space, 
compared to a $\chi^2$ distribution with one degree of freedom.
{\it Right panel:} the  spurious  ratio
$D_{fake}$ for $S_{cls}>4.0$, the white line shows the isolevel $D_{fake}=0.0$. 
}
\label{fig:signif}
\end{figure*}

\section{Cluster significance, background inhomogeneities, and  rejection of spurious clusters}
\label{sec:signif_sim}

Even though, we have identified the region of the $K$-\Eps~ parameter space,  
where the detection efficiency is larger,  and the probability to detect {\it fake} 
cluster is lower, in the application to real data, it's mandatory to 
provide a significance level, expressing the probability of  a 
cluster being not originated in a background fluctuation.
We propose a method derived   from the \cite{LiMa1983} approach, 
based on the evaluation of the  signal to noise (S/N) ratio. 
A significance method based on the S/N ratio fits well the the  
\gdbscan~ implementation, because the algorithm directly provides 
a partition of the photon list in {\it cluster} and {\it noise} 
events. Hence, for each  cluster we can evaluate easily the S/N 
ratio, knowing the exact nature of each event. The procedure to 
evaluate the significance is summarized by the following items:

\begin{figure*}
\centering
\includegraphics[width=18cm]{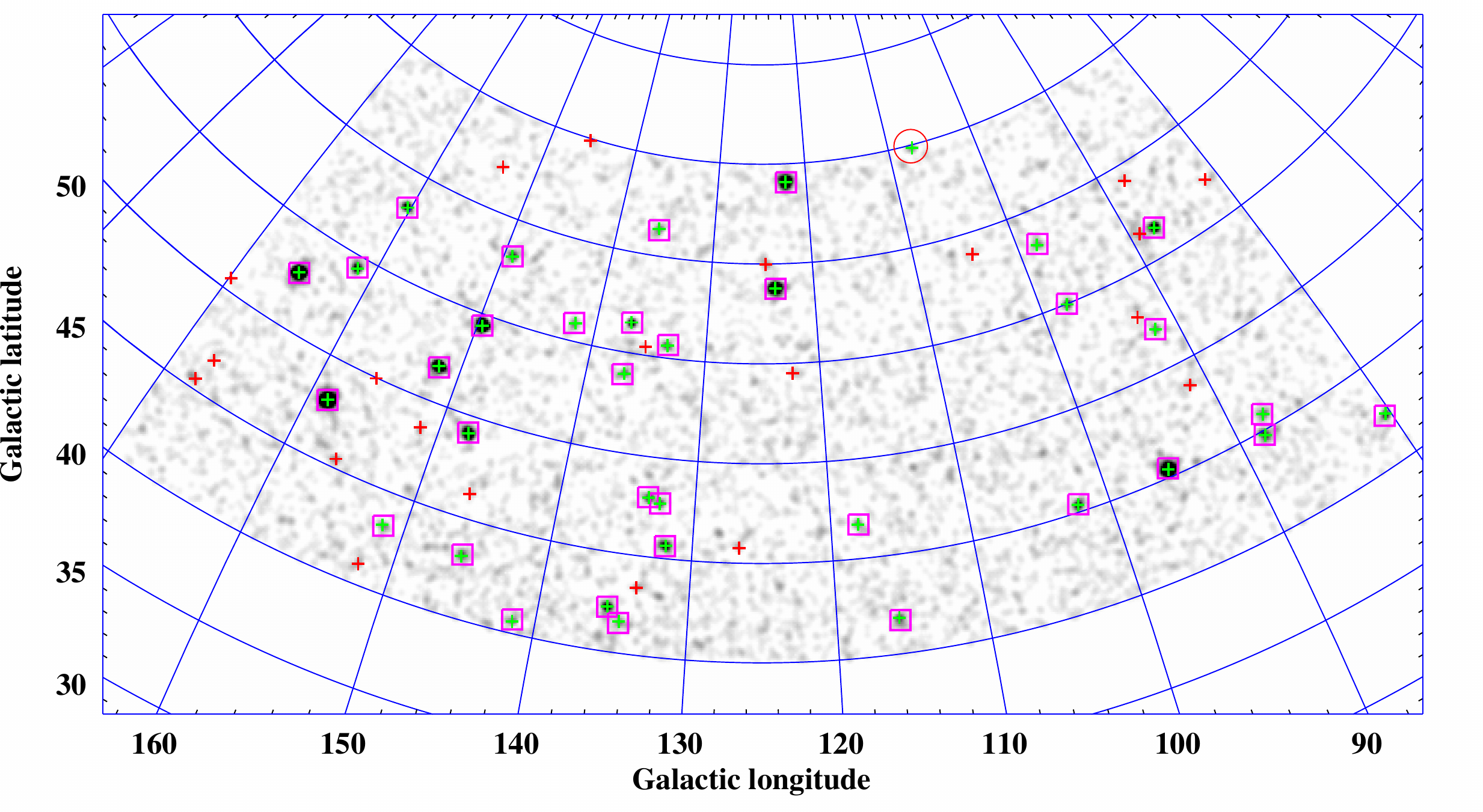}
\caption{Aitoff projection of the Fermi sky region. The purple boxes represent the \gdbscan~
 sources ($K=8,\varepsilon=0.21$ deg.). The green crosses are the 2FGL sources with $TS>16$, the
 red one those with $TS\leq16$. There are no {\it fake} sources, and the \gdbscan~ finds all the
 sources with $TS>16$, except only one, enclosed by the red circle, and with the center positioned 
 at the edge of the field.}
 \label{fig:Fermi_Cmap}
\end{figure*}

\begin{enumerate}
\item{} for each cluster, we define an annular region,  with 
an inner radius  $r_{in}$, and an external radius  $r_{out}$.
\item{} $r_{in}$ is set to an initial value of $r_{in}=2 r_{eff}$, and
 is adaptively increased with a step of $r_{in}/10$, for a
maximum of 10 trials, until at least the $95\%$ of the cluster events are enclosed within $r_{in}$.
\item{} $r_{out}$ is set to $3 r_{in}$.
\item{} We count all the cluster events $N^{in}_{src}$  and all the background events   
$N^{in}_{bkg}$, enclosed  within the circle with  radius $r_{in}$ and centered on the cluster  
centroid.  
\item{} We determine the $N^{out}_{bkg}$ background level, rescaling the number of 
background events in  $r_{in}<r<r_{out}$, to a circle with  radius $r_{in}$.  
\item{} To evaluate possible gradients in the background, we select a region enough far from the 
cluster to sample properly the background level, and enough close to the cluster, to measure 
a ”local” background level. At this regard, we define the radius $r_{out}^{ave}=(r_{out}+r_{in})/2$, 
and we evaluate the average background  level ($N^{local}_{bkg}$)  in a circle of radius $r_{in}$, centered on each point in  $r_{out}^{ave}<r<r_{out}$.
\item{} If no background points are { found} in  $r_{out}^{ave}<r<r_{out}$, we set 
$N^{local}_{bkg}=N^{out}_{bkg}$.  
\item {} By	comparing	$N^{local}_{bkg}$ to $N^{in}_{bkg}$,	we	evaluate the	fraction
of {\it noise} already resolved by the \gdbscan, and we evaluate the effective background level 
$N^{eff}_{bkg}$, by correcting $N^{local}_{bkg}$ for $N^{in}_{bkg}$.
\item{}  we evaluate the significance according to the Likelihood Ratio Test (LRT)
method proposed by \cite{LiMa1983}:

\begin{equation}
S_{cls}=\sqrt{2 \Big  ( N^{in}_{src} \ln \Big[ \frac{2 N^{in}_{src}}{N^{in}_{src}+N^{eff}_{bkg}} \Big]+
N^{eff}_{bkg} \ln \Big[ \frac{2N^{in}_{src}}{ N^{in}_{src}+N^{eff}_{bkg}} \Big ]  \Big  )}
\label{eq:signif}
\end{equation}

\end{enumerate}

\begin{figure}[t]
\centering
\includegraphics[width=9cm]{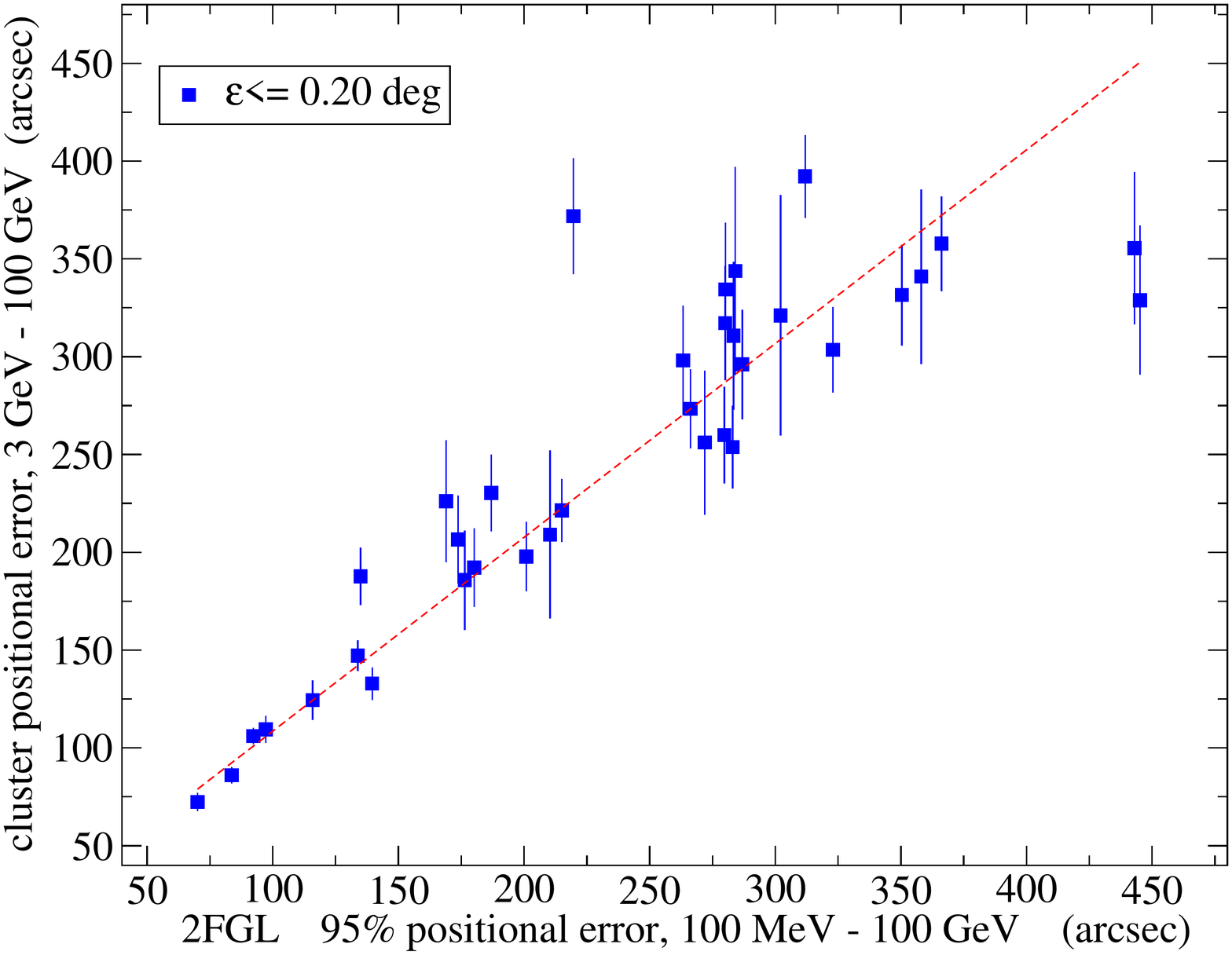}
\caption{
{
The scatter plot the positional error of the \gdbscan~ clusters vs. the 
positional error of the corresponding associated 2FGL$_{TS>16}$  sources. 
For each  2FGL$_{TS>16}$  source associated to one or more \gdbscan~ clusters, 
we plot the error on the position  of the reconstructed cluster centroid  and 
its standard deviation (represented by the error bar). 
The dashed red line represents a linear best fit with a slope
of $\simeq$ 0.99, and an intercept of $\simeq$ 9.53. 
}
}
\label{fig:Pos_rec_RealSky}
\end{figure}

Under the hypothesis that a cluster is due to a background fluctuation,
the variable $S_{cls}^2$, is expected to follow a chi square distribution,
with one degree of freedom ($\chi(1)^2$).  
In the left panel of Fig.  \ref{fig:signif}, we plot the  distribution of $S_{cls}^2$,
for the fake clusters in the {\it sky} test field 1  (blue histogram), 
compared to a  $\chi(1)^2$ distribution. 
The empirical distribution, is well described by the expected $\chi(1)^2$ distribution, proofing 
that the value of $S_{cls}$, can be used  as  the "significance" of the detected cluster. 
A very illustrative example of the power of $S_{cls}$ in rejecting {\it fake} clusters,  is  
given by the plot in the right panel in Fig \ref{fig:signif}, where we plot the
$D_{fake}$ ratio isolevel map, applying the selection  $S_{cls}>4.0$. { The {\it fake} ratio is 0 
for the parameter space with $\varepsilon\lesssim 0.25$ deg.  For 0.25 deg. $\lesssim \varepsilon \lesssim$ 0.35 deg.,  
there are { fluctuations} showing  $D_{fake}\lesssim 0.05$. Only for \Eps$\gtrsim 0.40$ deg. and 
$K\lesssim 8$, the  {\it fake} ratio shows a significant increase, but we stress that in this 
region of the parameter  space, \Eps~ is more then double of the PSF size, hence 
this is a region of the parameter space that should not be used in the detection with real data.
}

\section{Application to real \FLAT~ data }
\label{sec:realdata}

The last step in our investigation of the \gdbscan, is the application
to real \FLAT~ \gray~ data. We select the same region of the sky
used for the simulated test { field} ( $80^\circ<l<170^\circ$, and $40^\circ<b<65^\circ$),
and we extract all the photons with energy $E>3$ GeV. The photons
are collected for the same time span of the 2FGL
catalog . We repeat the detection 
test performed in the case of simulated data (see Sec. \ref{sec:det_sim} and  
Sec. \ref{sec:signif_sim}), restricting the parameter space to $2\leq K \leq10$,  
and $0.10\leq \varepsilon \leq0.30$ deg.

\begin{figure*}
\centering
\begin{tabular}{l} 
\includegraphics[width=17cm]{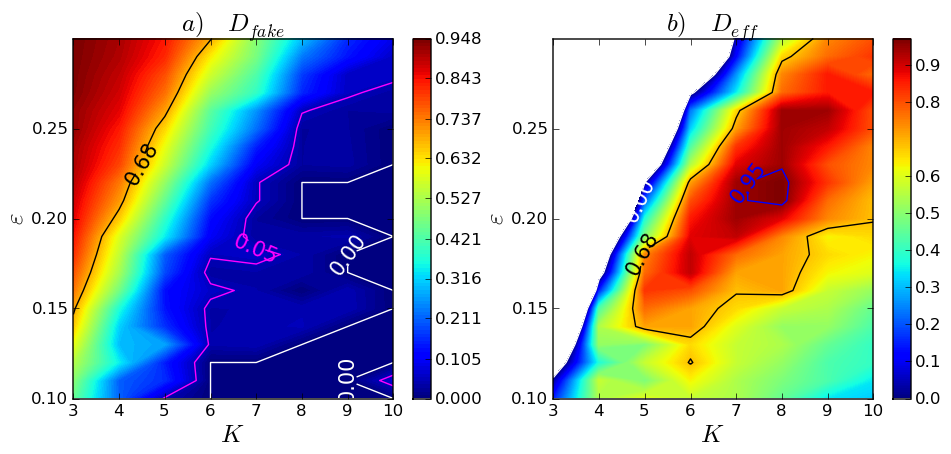}\\
\includegraphics[width=17cm]{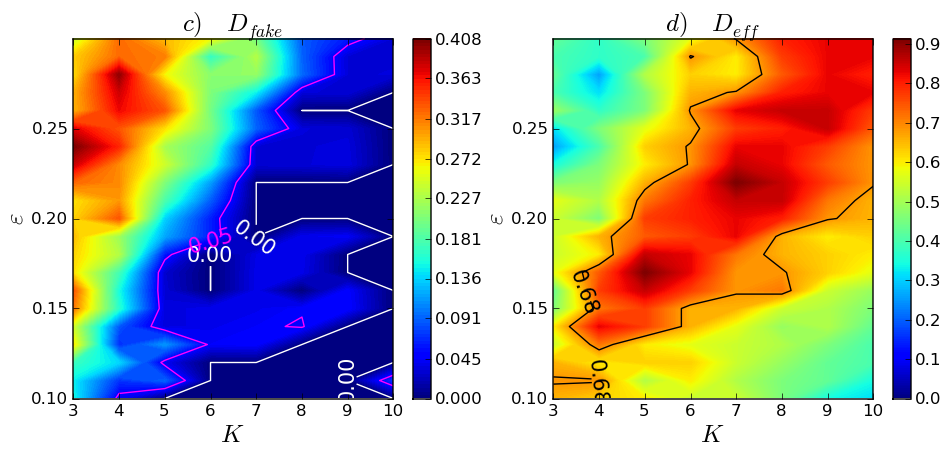}\\
\includegraphics[width=17cm]{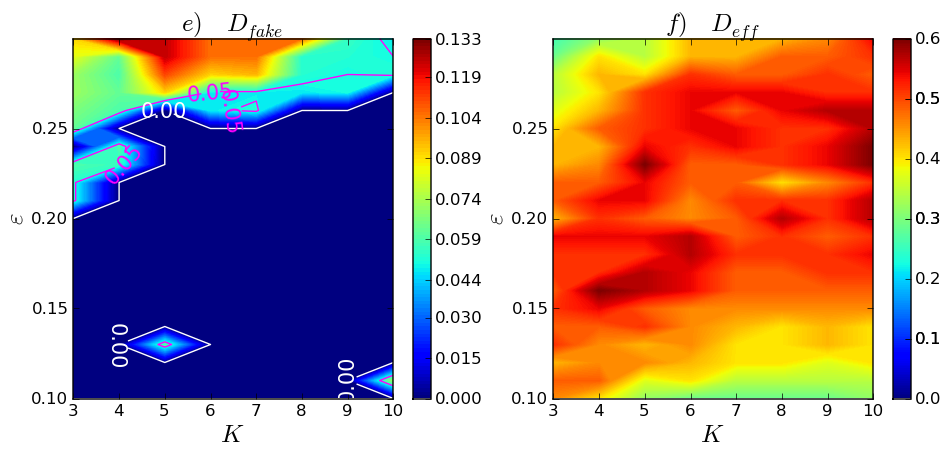}
\end{tabular}
\caption{
$D_{fake}$ (left panels) and $D_{eff}$ (right panels), for the
real sky detections, using the 2FGL$_{TS>16}$ catalog.
{\it Panels a,b:} no cut on $S_{cls}$ applied.
{\it Panels c,d:} $S_{cls}>2.0$
{\it Panels e,f:} $S_{cls}>4.0$ 
}
\label{fig:Fermi_Det}
\end{figure*}
\begin{figure*}[t]
\centering
\begin{tabular}{ll} 
\includegraphics[width=9cm]{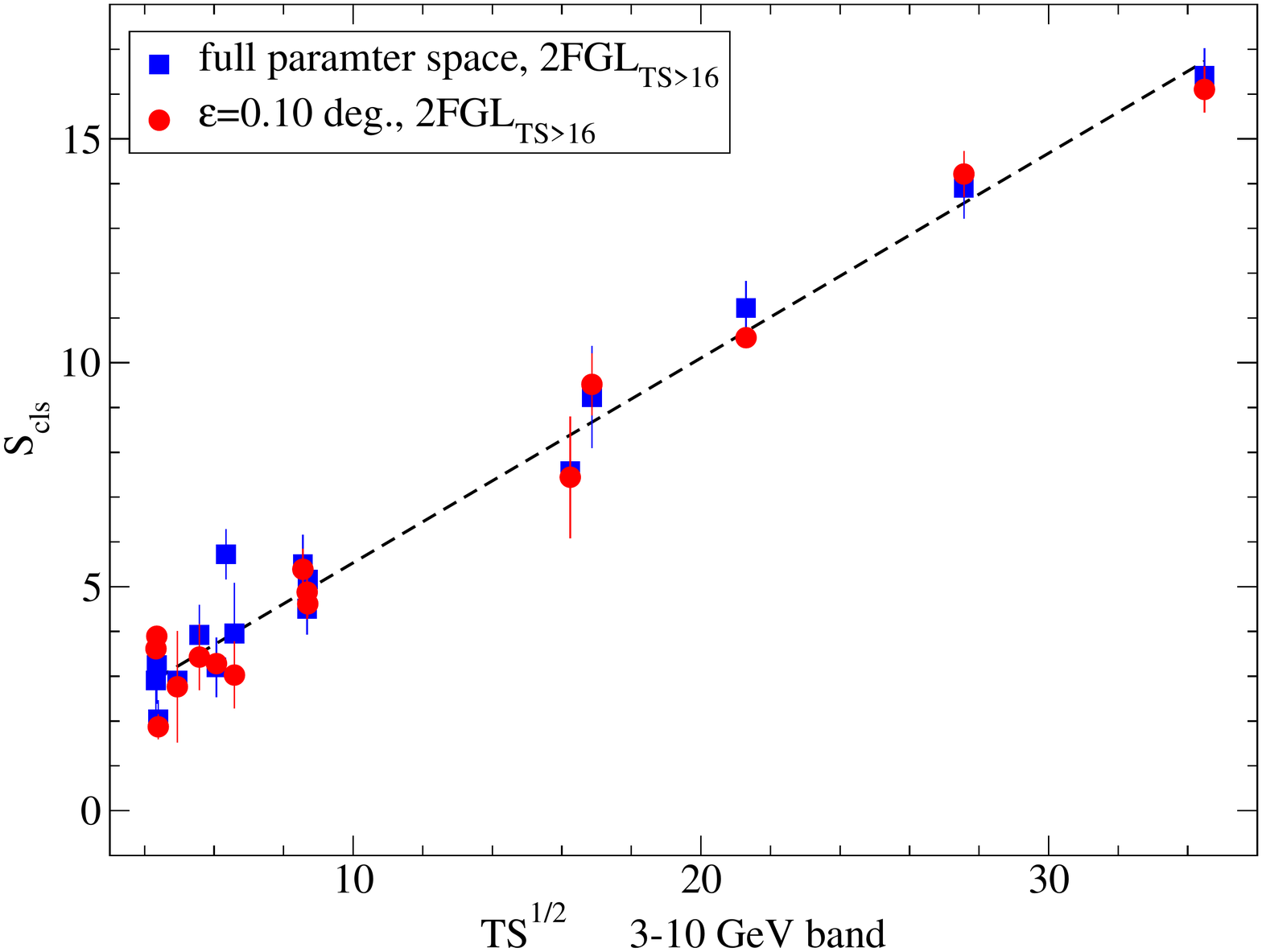}&
\includegraphics[width=9cm]{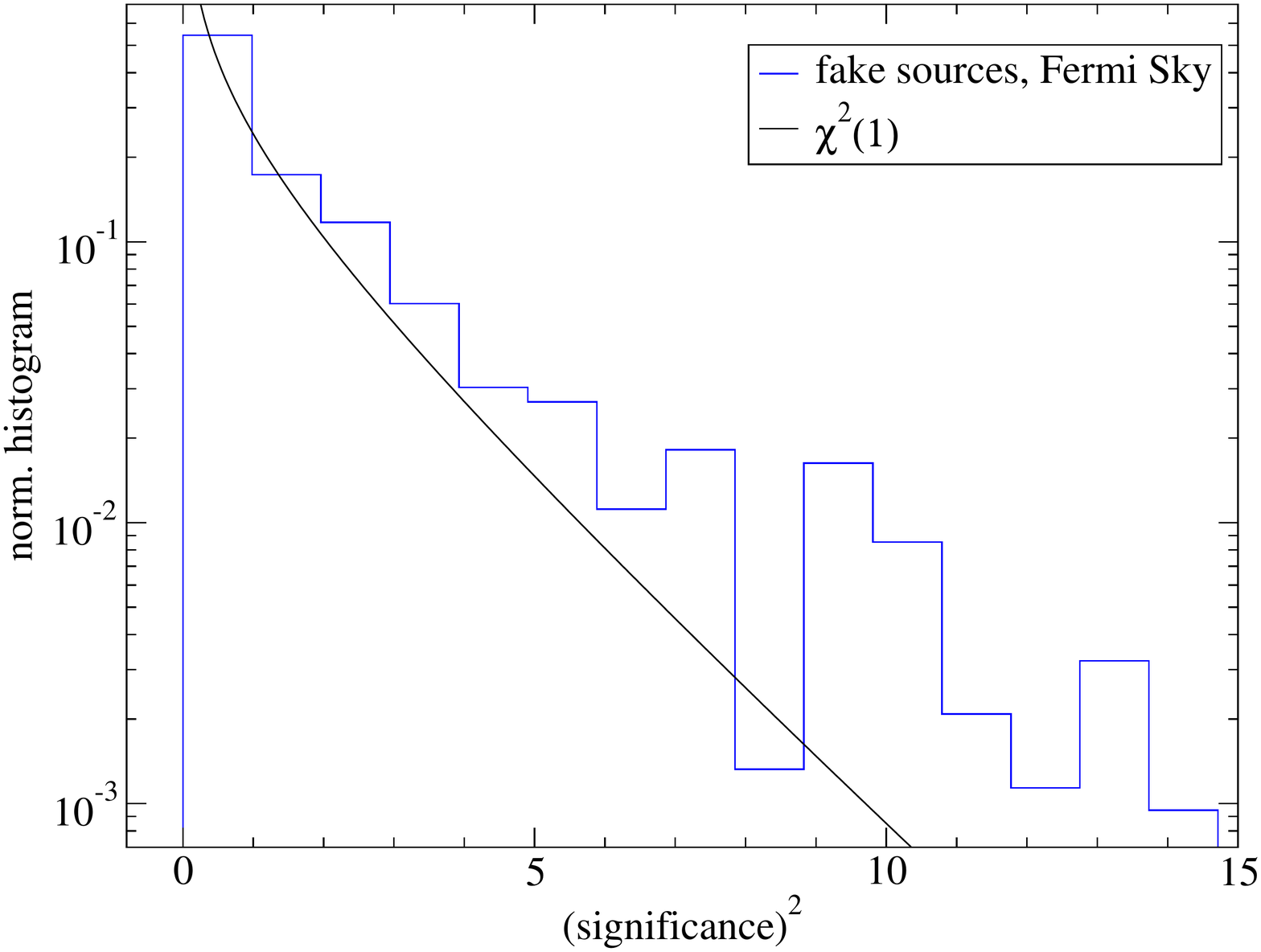}\\
\end{tabular}
\caption{
{\it Left panel: }scatter plot of $S_{cls}$ vs. $\sqrt{TS}$. For 
each source in our 2FGL$_{TS>16}$ list, associated to one or more \gdbscan~ cluster, we plot the  $\sqrt{TS}$ in the 3-10 GeV band, 
v.s. the average values of $S_{cls}$ and its standard deviation
(represented by the error bar).
{\it Right panel: }the distribution (blue line) of the square of the significance, for 
the {\it fake} clusters in the \FLAT~ real sky, for the full $K$,\Eps~ parameter space, 
compared to a $\chi^2$ distribution with one degree of freedom.
}
\label{fig:sig_vs_true_rate_RealSky}
\end{figure*}

To properly understand the detection performance, we need
to take into account that the 2FGL catalog has been built using
photons with an energy threshold of 100 MeV, whilst we use
a value of 3 GeV. A possibility is to select sources with
a reported flux larger than zero,  
in  the 3-10 GeV band flux column of the 2FGL.
This flux-based selection, is not the best way to 
study the detection performance of the \gdbscan~, indeed
the flux does not contain a unambiguous relation with
the significance of the detection, for that energy threshold. 
A more reliable  criterion is to select the sources according 
to the significance  reported in the 2FGL. The 2FGL detection 
significance is given by the $\sqrt{TS}$.  The $TS$ is the test 
statistic  defined as  $TS=2(\log L(\mathrm{source}) - \log L(\mathrm{no ~ source}) )$,
where $L$ is the likelihood of the data given the model with or without 
a source present at a given position on the sky \citep{2FGL}.
We apply a selection according to $\sqrt{TS}>4$, and we refer
to the corresponding source list (counting 35 sources) as
2FGL$_{TS>16}$.

An example of the application of the \gdbscan~ to real \FLAT~ 
data is given in Fig. \ref{fig:Fermi_Cmap}, where we report an Aitoff 
projection in galactic coordinates of the analysed \gray~ sky region. The red
crosses represent the 2FGL sources with $TS<16$ in the 3-10 GeV band, and the 
green crosses represents those with $TS\geq 16$. The purple boxes
represent the \gdbscan~ sources found for $K=8,~\varepsilon=0.21$ deg. 
For this choice of parameters, we find no {\it fake} sources,  and we find all 
the sources with with $TS>16$, except only one, enclosed by the red circle, and 
positioned at the edge of the sky region, with a galactic latitude $l=64.85$ deg.
In  Tab. 2 we summarize the detection performance, for  detections with a number  
of {\it fake}  sources $\leq$ 4.  We note that  values of {\it true} clusters ranges
between 35 and 34, out of the 35 present in the 2FGL$_{TS>16}$. 
The {\it fake} ones range between { 1} and 4, and we obtain an average detection
efficiency of  $D_{eff}= 0.94$.

In  Fig. \ref{fig:Pos_rec_RealSky} we compare the localization 
performance of the \gdbscan~ algorithm  with that returned by the 
likelihood analysis implemented in the Fermi Science Tools. 
For each source in our 2FGL$_{TS>16}$ list, 
associated to one or more \gdbscan~ clusters, we plot the  the error on the position 
of the reconstructed cluster centroid  and its standard deviation (represented by the error bar) 
vs. the 95$\%$ positional  uncertainty reported in the 2FGL. We evaluate the
2FGL  95$\%$ positional  uncertainty as $\sqrt{\sigma_{95,min}~\sigma_{95,max}}$,
where $\sigma_{95,min}$ and $\sigma_{95,max}$, are the semimajor and semiminor axes of the 
95$\%$ confidence source location region, respectively.
The dashed red line represents a linear best fit, with a slope
of $\simeq$ 0.99, and an intercept of $\simeq$ 9.53, showing that 
the error on the position of the reconstructed cluster centroid, performed with a threshold of 3 GeV, is 
of the same order of the $95\%$ positional uncertainty reported in the 2FGL catalog, performed 
above 100 MeV.

To test the reliability of  the significance $S_{cls}$ to reject spurious sources, 
 in Fig. \ref{fig:Fermi_Det}  we plot the $D_{fake}$ and $D_{eff}$, based
on the 2FGL$_{TS>16}$ catalog. The panels a and b,  correspond to the case of
no selection on $S_{cls}$. Both the $D_{fake}$  and the $D_{eff}$ trends are 
very similar to the case of the simulated sky.
If we apply a significance cut of $S_{ls}>2.0$ (panels c,d), we observe that
the number of spurious ratio is $D_{fake}\leq$ 0.05 for  almost half of the parameter space
(region to the right of the purple line). 
 The more severe cut of $S_{cls}>4.0$
(panels d,e), removes  all the {\it fake} clusters except two, { for 
$\varepsilon \lesssim$ 0.15 deg. Only for  $\varepsilon \gtrsim$ 0.25 deg., the $D_{fake}$ 
ratio shows a significant increase, ranging from 0.05  up to $\simeq$ 0.1.  
In agreement with our analysis on simulated data, the region of the parameter space 
where \Eps~ is comparable to the PSF size, gives the better performance.
} 

To have a further confirmation about the robustness of our significance,
we plot in the right panel of Fig. \ref{fig:sig_vs_true_rate_RealSky},
$S_{cls}$ vs. $\sqrt{TS}$. For each source in our 2FGL$_{TS>16}$ list, 
associated to one or more \gdbscan~ cluster, we plot the  $\sqrt{TS}$ 
in the 3-10 GeV band, 
v.s. the average value of $S_{cls}$ and its standard deviation
(represented by the error bar). The average value of $S_{cls}$  and its
standard deviation  are evaluated from  the list of all the cluster associated
to  the same  2FGL source. The solid blue boxes represent the full $K$,\Eps~
parameter space case, and the red solid circles represent the $\varepsilon=0.10$ deg. case.
The dashed black line represents a linear best fit. 
The slope of the linear fit is $\simeq 0.5$. The strong correlation in
the scatter plots ($r\simeq 0.98$, for both data sets), proves that  our
significance implementation is consistent with the $\sqrt{TS}$  reported
in the 2FGL, and the slope of the linear fit suggests that $S_{cls}\simeq 0.5 \sqrt{TS}$.

\begin{table}

\begin{tabular}{@{}l@{}|@{}c@{}ccccccc@{}}
\hline
     & 2FGL$_{TS>16}$ & True & Fake & $K$ & \Eps~ & $D_{eff}$ & $Q$  \\ 
\hline
\hline 
    
{Fermi-{\it}sky}   & 35& 34& 0& 8& 0.21& 0.97& 0.97& \\ 
                          & 35& 34& 1& 7& 0.20& 0.94& 0.92& \\ 
                          & 35& 35& 2& 7& 0.21& 0.94& 0.89& \\ 
                          & 35& 35& 4& 6& 0.19& 0.89& 0.79& \\ 
\hline
average  &  35& 34.50& 1.75& 7.00& 0.20& 0.94& 0.89& \\ 
\hline
\hline

\hline  
\end{tabular}

\label{tab:tab2}
\caption{Summary of the detection performance for the
real \FLAT~  field, for detections with a number of fake sources $\leq 4$.
}
\end{table}
\section{Conclusions}
\label{sec:conclusions}
For the first time, we have used the \dbscan~ for the detection of sources 
in \gray~  astrophysical images. We have implemented a new version of
the  \dbscan, the \gdbscan, that is optimized for the application to \gray~   
astrophysical images, with relevant background noise. Our \gdbscan, presents
the novelty of recursive call of the \dbscan~ algorithm, that allows an
excellent reconstruction of the cluster, with an effective background
rejection.
We have tested the algorithm with a sample of simulated \gray~\FLAT~ fields,
to give a statistical characterization of the method, and to benchmark
the detection performance. The results, with the simulated \gray~ data,
are summarized by the following items:
\begin{itemize}
\item{} The radius of \gdbscan~ scanning brush \Eps~, has a strong 
correlation with the instrumental PSF radius. We find that 
the typical size of the reconstructed {\it true} cluster is of the
order of the simulated  PSF size  $\sigma^{sim}$,
{ and that the precision of the reconstructed
centroid is of the order of  $\sigma^{sim}/\sqrt{N_p}$. 
}
\item{} The number of reconstructed events $N_p$ is ruled by the 
Poissonian statistics in the {\it random } fields, and for
the {\it fake} clusters. On the contrary, for {\it true} clusters,
the statistics of $N_p$, is ruled by that of the simulated sources.
\item{} The fractional error  on the reconstructed events number is 
of the order of $20\%$ for  $N_p sim. \lesssim 50$, and is negligible
for larger values, with best performance obtained when 
\Eps$\simeq \sigma^{sim}$.
\item{} We have investigated the detection performance, for 
a wide range of the $K$,\Eps~ parameter space, and we have
identified the region with the best performance in terms
of detection efficiency, and spurious ratio.
\item{} We have implemented an algorithm for the estimate 
of the Signal to Noise (S/N) ratio, able to deal with local
background inhomogeneities and nearby sources contamination, 
and we have successfully used the S/N estimate to determine 
the significance of the clusters, using the definition in \cite{LiMa1983}.
\item{} Our cluster significance, $S_{cls}$, for random clusters,
follows the $\chi(1)^2$ statistics, and can be used to
reject spurious sources. The chance to find spurious
sources for $S_{cls}>4$, is negligible. This means, that
our $S_{cls}$ is a robust a reliable tool to reject spurious sources,
and that $\chi(1)^2$ statistics can be used to evaluate the probability of
a cluster to be spurious.
\end{itemize}

We have successfully applied the \gdbscan~ to real \FLAT~ data.
We have found an excellent agreement with  results from the simulated fields.
We tested our detection performance using as catalog, the  2FGL
sourced with a $\sqrt(TS)>4$ cut.
The results, with the real \FLAT~ \gray~ data,
are summarized by the following items:
\begin{itemize}
{
\item{} the error on the position of the reconstructed cluster centroid, 
performed with a threshold of 3 GeV, is of the same order of the 95$\%$ 
positional uncertainty reported in the 2FGL,  performed above 100 MeV. 
}
\item{} We tested the \gdbscan~ significance, finding that it is
strongly correlated with the $TS$ provided in the 2FGL.
The significance cut, allows to remove safely spurious 
clusters.
\item{} The detection efficiency with real data is excellent,
we are able to find all the 35 sources with $\sqrt(TS)>4$.
\item{}  When working with \Eps~ of the order of the instrumental
PSF size, we obtain the best performance, in terms of spurious 
rejection, and detection efficiency 
\end{itemize}

In general, we find that the \gdbscan~ is a very powerful detection method
to find clusters in \gray~ images, corresponding to real sources.
It has the great advantage to deal self-consistently  with
gradient in the background, providing an effective rejection of spurious
clusters. Our implementation of the detection significance, in addition
to the algorithm to evaluate local fluctuations in the background,
allows to apply statistically significant selection, making even
more effective the rejection of spurious sources.

In a companion paper \citep{dbscan_inprep}, we will a apply the method
to the \FLAT~ sky, showing the potentiality for the discovery of new sources,
in particular of small clusters located at high galactic latitude, or
cluster on the galactic plane, affected by a strong background.
{
We will also investigate how to plug the energy dependence
of the PSF into the \gdbscan~ algorithm, and how to improve
the detection performance taking into account other 
\FLAT~ calibration properties.
}

We  remark that, since the \gdbscan~ provides also density maps,
it can potentially be used in the detection of large  scale structures 
in the galactic \gray~ background, providing patterns to compare to the
interstellar gas distribution.  
We also stress, that the application of this method are not limited 
to \gray~ images, but can be potentially used for any application
related to the detection of spatial, and/or spatio/temporal clusters.

\begin{acknowledgements}
We are grateful to Enrico Massaro, Riccardo Camapana, and Enrico Bernieri, 
for helpful comments, and for providing us the simulated test fields.
We are grateful to Gino Tosti, for helpful comments.
We thank the anonymous referee for providing us with constructive comments and useful suggestions.
\end{acknowledgements}

\bibliographystyle{aa} %
\bibliography{dbscan}

\end{document}